\documentclass{article}
            \newcommand{\ad}{a^{\dag}}
            \newcommand{\cd}{c^{\dag}}
            \newcommand{\dd}{d^{\dag}}
            \newcommand{\ads}{(a^{\dag})^{s}}
            \newcommand{\adt}{(a^{\dag})^{t}}
            \newcommand{\p}{\prime}
            
            \newcommand{\spp}{s^{\p\p}}
            \newcommand{\tp}{t^{\p}}
            \newcommand{\g}{\gamma}
            \newcommand{\gp}{\gamma^{\p}}
            \newcommand{\gpp}{\gamma^{\p\p}}
            \newcommand{\asat}{h,\g,s;h_{1},\delta,t}
            \newcommand{\asatp}{h',\g',s';h'_{1},\delta',t'}
            \newcommand{\asatpp}{h'',\g'',s'';h''_{1},\delta'',t''}
            \newtheorem{theorem}{Theorem}

            \usepackage{amsfonts}
            \usepackage{graphicx}
            
\begin{document}
           \title{Reference Frame Fields based on Quantum
           Theory Representations of Real and Complex Numbers}
          \author{Paul Benioff,\\
          Physics Division, Argonne National
           Laboratory,\\ Argonne, IL 60439, USA \\
           e-mail:pbenioff@anl.gov}

           \maketitle

           \begin{abstract}
             A quantum theory representations of real (R) and complex
            (C) numbers is given  that is based on states of
            single, finite strings of qukits for any base $k\geq 2.$
            Arithmetic and transformation properties of these states
            are given, both for basis states representing rational
            numbers and linear superpositions of these states.
            Both unary representations and the possibility that qukits
            with $k$ a prime number are elementary and the rest
            composite are discussed. Cauchy sequences of $q_{k}$ string
            states are defined from the arithmetic properties. The
            representations of $R$ and $C,$  as equivalence classes of
            these sequences, differ from classical representations as
            kit string states in two ways: the freedom of choice of basis
            states, and the fact that each quantum theory representation
            is part of a mathematical structure that is itself based on the real
            and complex numbers. In particular, states of qukit strings
            are elements of Hilbert spaces, which are vector spaces over
            the complex field. These aspects enable the description of
            $3$ dimensional frame fields labeled by different k values,
            different basis or gauge choices, and different iteration stages.
            The reference frames in the field are based on each R and C
            representation where each frame contains representations of all
            physical theories as mathematical structures based on the
            R and C representation. Some approaches to integrating this
            work with physics are described. It is observed that R and C
            values of physical quantities, matrix elements, etc. which are
            viewed in a frame as elementary and featureless, are seen in a parent
            frame as equivalence classes of Cauchy sequences of states of
            qukit strings.
           \end{abstract}

           \section{Introduction}
           Numbers play a basic role in physics and mathematics, so basic in fact
           that their use, both in experiments and in theory, is
           taken for granted and is rarely examined. Natural numbers
           and integers are probably the most basic because of their role in
           counting, rational numbers play a basic role in that
           numerical experimental outputs are represented as
           rational numbers.  They also are the type of numbers used
           in all computer computations.

           The importance of real and complex numbers lies in their
           being the number base of all physical theories used so
           far. This includes classical and quantum mechanics,
           quantum field theory, QED, QCD, string theory, and
           special and general relativity. Each of these theories is
           a mathematical theory
           characterized by a different set of axioms. Assuming the
           axiom sets are consistent, each theory has many different
           representations as mathematical structures based on the
           real and complex numbers.  The connection to physics is
           made by interpreting some of the elements in the
           mathematical structures as representing physical systems
           and physical quantities. Examples include the use, in quantum
           theory, of elements of Hilbert spaces and operators on the
           spaces to represent states and observable physical quantities of
           systems, the use of other elements,  to represent various
           properties of space time, etc.

           In all of this, the tacit assumption is made that the
           properties of physical systems and the physical universe
           are independent of the properties of mathematical
           theories and their representations.  The general approach
           taken is to discover the theory that best describes
           physical systems and their properties. Little attention is
           paid to whether the basic properties of theories and
           their mathematical representations have any influence on the
           basic properties of physical systems or how intertwined
           physics and mathematics are.

           The approach taken in this paper stems from the work of
           Wigner on the unreasonable effectiveness of mathematics in the
           natural sciences \cite{Wigner,Hamming,Boniolo}. One
           answer to this problem is that one should work towards
           developing a coherent theory of mathematics and physics together
           \cite{BenTCTPM,BenTCTPMTEC}. Presumably such a theory
           would show why mathematics is important to physics.

           This paper is, hopefully, a step in this direction. Here
           extension of previous work on the quantum representation
           of numbers \cite{BenRRCNQT,BenFIQRF} shows
           that quantum theory representations of real and complex
           numbers have properties not possessed by classical
           representations of these numbers.  It will be seen that
           the structures resulting from these properties suggest a
           close intertwining between the properties of physical and
           mathematical systems.

           Although little investigated, these possibilities are not
           new.  Perhaps the closest is the work of Tegmark
           \cite{Tegmark1,Tegmark2} which suggests that the
           physical universe really is a mathematical structure.
           Other work which emphasizes the close
           relationship between physics and mathematics is concerned
           with quantum theory representations of mathematical
           systems.  This work includes papers on quantum set theory
           \cite{Finkelstein,Schlesinger,Titani}, quantum theory
           representations of real numbers
           \cite{Takeuti,Ozawa,Tokuo,Corbett,Davis,Gordon,Litvinov},
           and the use of category theory in physics
           \cite{Isham,Krol}.

           The quantum representations of real and complex numbers
           presented here differ from other work in this area in that
           they are not abstract representations based on quantum
           logic or on lattice valued models of set theory
           \cite{Takeuti,Ozawa,Tokuo,Davis,Gordon}, nonstandard
           numbers \cite{Corbett}, or category theory \cite{Isham,Krol}.
           Instead they are based on representations of natural numbers, $N$,
           integers, $I$, and rational numbers, $Ra$, as states of finite
           strings of qukits.\footnote{Qukits are extensions of qubits to
           systems with states in a finite $k$ dimensional Hilbert space.}

           This choice is based on the observation that all physical
           representations of numbers are in the form of k-ary
           representations as states of strings of kits or of
           qukits. This is the case for all experimental outputs.
           Also all computations are based on these representations
           of numbers.  The importance of this type of number for
           computations and the limits of computation suggest other ties
           to information theory and limitations on the information
           resources of the  universe \cite{Lloyd,Davies,Hsu}. The
           restriction here to qukit strings is based on the fact
           that quantum theory is the basic underlying theory of all
           physical systems.

           Here the quantum theory representations of real numbers are
           described as equivalence classes of Cauchy sequences of
           states of qukit strings. In essence this is a translation
           of the definition in mathematical analysis textbooks
           \cite{Hewitt,Burkhill} as equivalence classes of Cauchy sequences
           of rational numbers into quantum theory.\footnote{An often
           used equivalent definition is based on Dedekind cuts of
           rational numbers instead of Cauchy sequences.}

           These representations are described in the next two
           sections. First quantum representations of natural numbers,
           integers, and rational numbers are presented as states of
           single finite qukit strings. These are based on the states
           of each qukit as elements of a $k$ dimensional Hilbert space.
           These are used in the quantum representations
           of real numbers as equivalence classes of Cauchy sequences of
           states of single finite qukit strings.

           Quantum representations of real and complex numbers
           differ from classical representations in several ways.
           One  difference is that the equivalence classes of Cauchy
           sequences of qukit string states are larger than
           classical classes as they contain sequences that do not
           correspond to any classical sequence.  However, no new
           equivalence classes are created.

           A more important difference is that, for states of qukit
           strings, there is a freedom of basis state choice that
           does not exist in classical representations. This is
           based on the observation that the states of each qukit
           are elements of a $k$ dimensional Hilbert space. In order
           that states of qukits, $(q_{k}),$ represent numbers, one
           must choose a basis set of states for each $q_{k}$ in the
           string. This is well known in quantum computation where
           binary representations of numbers, such as
           $|1100101\rangle$ as a state of a qubit string, imply a
           choice of basis for each qubit. This freedom of basis
           choice is also referred to here as a gauge freedom or
           freedom to fix a gauge for each $q_{k}$.  It is
           represented here by a variable $g$ that ranges over all
           basis  or gauge choices for $q_{k}$ states in a string.
           This gauge freedom is seen to extend up to
           representations of real numbers in that for
           each gauge choice $g$ one has quantum theory
           representations $R_{k,g}$ of real
           numbers that are different for different $k$ and $g$.
           Even though these representations are $k,g$ dependent,
           they are all isomorphic to one another.

           These representations for different $k$ and $g$ are
           described in section \ref{SRNRAT}. Both base changing
           transformations and gauge transformations are
           described for the finite $q_{k}$ string states. Lifting
           these up to transformations on the Cauchy sequences gives
           transformations that take one real number representation
           to another, $R_{k,g}\rightarrow R_{k',g'}.$

           The description is extended to include quantum theory
           representations of complex numbers, $C_{k,g},$ in section \ref{QRCN}.
           They are defined as equivalence classes of Cauchy
           sequences of states of pairs of finite $q_{k}$ strings
           where the pair elements correspond to real and
           imaginary parts of a complex rational number.
           Cauchy conditions are applied separately to the sequences
           of real and imaginary components.

           There is another very important difference between
           quantum and classical representations of real and complex
           numbers.  This is the fact that the states of the $q_{k}$
           strings used to define Cauchy sequences are elements of a
           Fock space that is itself a vector space over a field of
           real and complex numbers. For example all eigenvalues of
           operators acting on these string states are complex or
           real numbers.  Also all  linear superposition
           coefficients are complex numbers. This is quite different
           from the classical situation in that real and complex
           numbers play no role in the representation of numbers as
           states of bit or kit strings.

           This dependence of quantum theory representations on the
           real and complex number base of spaces of $q_{k}$ string
           states leads to the possibility of iteration of the
           construction. Each representation $R_{k,g},C_{k,g}$ can
           serve as the real and complex number base of Hilbert
           space and Fock space representations of $q_{k}$ string
           states that can be used to construct other
           representations of the real and complex numbers.

           In addition, this same iteration possibility extends to
           all physical theories that are representable as
           mathematical structures over the real and complex
           numbers. Included are quantum and classical mechanics,
           quantum field theory, special and general relativity,
           string theory, as well as other theories.

           This leads to the association of a reference frame
           $F_{k,g}$ to each representation $R_{k,g},C_{k,g}.$ Each
           frame $F_{k,g}$ contains representations of all physical
           theories as mathematical structures based on $R_{k,g},C_{k,g}.$
           This use of reference frame terminology is consistent with
           other uses \cite{Aharonov,Bartlett} in that it sets a base or
           reference point $R_{k,U},C_{k,U}$ for representations
           of all physical theories.

           Much of the rest of the paper, Section \ref{FQRF}, is
           concerned with properties of these reference frames and
           with three dimensional fields of these reference frames.
           Two of the dimensions are labeled by $k$ and $g$. The
           third is by an integer $j$ denoting the iteration stage.
           Different iteration possibilities are considered:, finite,
           one way infinite, two way infinite, and cyclic. Also
           properties of observers in different locations in the
           frame field are described.

           Section \ref{IP} includes a discussion on what is probably
           the most important outstanding issue, how to integrate the
           frame field with physics. This is especially important
           from the viewpoint of constructing a coherent theory of
           physics and mathematics together \cite{BenTCTPM,BenTCTPMTEC}
           or if one considers the physical universe as a mathematical
           universe \cite{Tegmark1,Tegmark2}. Both relatively simple
           aspects of the possible integration, and more speculative
           aspects are described.  However it is clear from this that
           much remains to be done to achieve an integration with physics.

           The discussion section includes a description of the possible
           replacement of Cauchy sequences by operators, a possible use
           of gauge theory to integrate this work with physics, and
           other issues.

           Two aspects of the following work should be emphasized.
           One is that rational numbers are represented
           by states of \emph{single} qubit strings and not by
           states of pairs of qubit strings. This is based on the
           observation that all physical representations of rational
           numbers, such as computer inputs and outputs, outcomes of
           measurements, physical constants, etc.  are as single
           strings of digits in some base $k\geq 2$ (usually $2$ or $10$)
           and not as integer pairs.  Also complex numbers are
           represented in computations by pairs of finite digit strings
           in some base where the pair elements correspond to the real
           and imaginary components. The use of this representation
           is based on the fact that  sets of rational numbers so
           represented are dense in the sets of all rational and
           all real numbers.

           In this paper basic arithmetic relations and operations
           for the different types of numbers are discussed. The reason
           for this is based on the observation
           that states of kit or qukit strings, such as $|100101\rangle$
           for $k=2$, do not, in any ab initio sense, represent numbers
           of any type. In order to show that these states represent numbers,
           one must prove that they satisfy a relevant set of
           axioms.  The axioms are expressed in terms of
           properties of basic arithmetic relations and operations.
           It follows that a proof that sets of states of finite
           $q_{k}$ strings represent numbers is based on showing
           that definitions of these relations and operations satisfy
           the relevant axiom sets. Some details of these
           proofs, which are based on classical proofs, \cite{Hewitt},
           are given in \cite{BenRRCNQT,BenSQTRN}.

           \section{Quantum Representations of Natural Numbers,
           Integers, and Rational Numbers}\label{QRNNIRN}

           \subsection{Representations}
           The quantum representations of numbers are described here
           by states of  strings of qukits on a two dimensional integer
           lattice, $I\times I$. The states are given by $|\g,0,h,s\rangle_{k,g}$
           where $s$ is a $0,1,\cdots,k-1$ valued function on an interval
           $[l,h;u,h]$ of $I\times I,$ with $l\leq 0\leq u,$ $\g = +,-$
           denotes the sign, and $0,h$ the lattice location of the $k-al$
           point. The reason for the subscript $g$ will be clarified
           later on.

            Here it is intended that the states $|\g,0,h,s\rangle_{k,g}$
           represent numbers in $N,I,$ and $Ra.$ For numbers in $N,$
           $\g=+,l=0$; for numbers in $I$, $l=0,$  and there are
           no restrictions for $Ra.$ A compact notation is used where
           the location of the sign, denoted by $0,h$, is also the
           location of the $k-al$ point. As examples, the base $10$ numbers
           $612,\; -0474,\; -012.7100$ are represented here by
           $|612+\rangle,\; |0474-\rangle,\; |012-7100\rangle$
           respectively. Note that leading and trailing $0s$ are
           allowed.

           The states $|\g,0,h,s\rangle_{k,g}$ can be represented in terms of
           creation operators acting on the qukit vacuum state
           $|0\rangle$  where \begin{equation}\label{g0hs}
           |\g,0,h,s\rangle_{k}=\cd_{\g,0,h}\ad_{s(u,h),u,h}\cdots
           \ad_{s(l,h),l,h}|0\rangle =\cd_{\g,0,h}(\ad)^{s}_{h}|0\rangle.
           \end{equation} Here  $\cd_{\g,0,h}$ creates a sign qubit at
           $(0,h)$ and $\ad_{i,j,h}$ creates a qukit in state
           $i=0,1,\cdots,k-1$ at $ (j,h).$  $(\ad)^{s}_{h}$ is a
           short representation of the string of $\ad$ operators.

           The creation operators and the corresponding
           annihilation operators satisfy the usual commutation or
           anticommutation rules for respective boson or fermion qukits.
           The variable $h$ is present to allow for the presence of
           $n-tuples$ of $q_{k}$ string states representing $n-tuples$
           of numbers.

           The use of $I\times I$ as a framework for qukit state
           representations is based on the need to distinguish qukits in
           a string by a discrete ordering parameter and to
           distinguish different qukit strings from one another. This
           is seen in Eq. \ref{g0hs} where the integers $j$ with
           $l\leq j\leq u$ order the qukits in a string and the
           values of $h$ serve to distinguish different strings.
           There is no need to consider $I\times I$ as a lattice
           of points in a two dimensional physical space
           as its sole function is to provide discrete
           ordering and distinguishing labels.

           Also the locations of the qukit strings in the lattice
           direction of the strings will be restricted in that the
           sign qubit will always be at site $0$. This
           restriction is inessential because the only function
           of the $j$ label in $(j,h)$ is to provide a discrete
           ordering of qukits states in a string.

           The set of states $|\g,0,h,s\rangle_{k,g}$ for all $\g,h,s$
           are a basis, $\mathcal{B}_{k,h,g},$ that spans a Fock space
           $\mathcal{F}_{k,h}$ of states that are
           linear superpositions \begin{equation}\label{linsuppsi}
           \psi=\sum_{\g,h,s}c_{\g,h,s}|\g,0,h,s\rangle_{k,g}\end{equation}
           Here and in the following, $\sum_{s}=\sum_{l\leq 0}
           \sum_{u\geq 0}\sum_{s_{[l,u]}}$ is a sum over all integer
           intervals $[l,u]$ and over all $0,1,\cdots,k-1$ valued
           functions $s$ with domain $[l,u].$ A Fock space is used
           because states of $q_{k}$ strings with different numbers of
           qukits are included. The subscript $k\geq 2$  denotes the
           base.  Note that base $k$ qukits are different from base
           $k'$ qukits just as spin $k$ systems are different from
           spin $k'$ systems.

           Extension of the description to include pairs, triples and
           $n-tuples$ of basis states and their linear superpositions is
           done by distinguishing different states in the tuples with different
           values of $h$. For each finite subset $S =h_{1},h_{2},\cdots,
           h_{|S|}$ of integers where $|S|$ is the number of integers in $S$,
           let $\mathcal{B}_{k,S,g}$ be the set of states of the form
           $|\g_{1},0,h_{1}s_{1}\rangle_{k}|\g_{2},0,h_{2},s_{s}
           \rangle_{k}\cdots|\g_{|S|},0,h_{|S|},s_{|S|}\rangle_{k}.$
           Define $\mathcal{B}_{k,g}$ by\begin{equation}\label{basiskg}
           \mathcal{B}_{k,g}=\bigcup_{S}\mathcal{B}_{k,S,g}.\end{equation}
           $\mathcal{B}_{k,g}$ is a basis set of all finite tuples of
           states of finite $q_{k}$ strings. Let $\mathcal{F}_{k}$
           be the Fock space spanned by the states in $\mathcal{B}_{k,g}.$

           The representation of state $n-tuples$ used here is by
           products of states as in $|\g_{1},0,h_{1},s_{1}\rangle_{k,g}\cdots
           |\g_{|S|},0,h_{|S|},s_{|S|}\rangle_{k,g}.$  The A-C operator
           representation of this state is $\cd_{\g_{1},0,h_{1}}
           (\ad)^{s_{1}}_{h_{1}}\cdots\cd_{\g_{|S|},0,h_{|S|}}
           (\ad)^{s_{|S|}}_{h_{|S|}}|0\rangle.$ For bosons the
           ordering of the operators is immaterial.  For fermions a
           specific ordering must be selected as a canonical
           ordering.

           The basic arithmetic relations needed to show that the states
           $|\g,0,h,s\rangle_{k,g}$ do represent numbers are equality
           $=_{A,k,g}$ and less than $<_{A,k,g}$.\footnote{One cannot
           avoid defining these relations and operations directly on
           the states. To see this let the operator $\tilde{N}$ satisfy
           $\tilde{N}|\g,0,h,s\rangle_{k,g}=N(\g,s)|\g,0,h,s\rangle_{k,g}$
           where $N(\g,s)$ is supposed to be the  number
           represented by $|\g,0,h,s\rangle_{k,g}.$
           Because of the possible presence of leading and trailing
           $0s$, the eigenspaces of $\tilde{N}$ are infinite
           dimensional. One knows that the set of eigenvalues of
           $\tilde{N}$ satisfy the relevant axioms. To prove
           that $N(\g,s)$ is the  number represented by
           $|\g,0,h,s\rangle_{k,g}$ one must show that $\tilde{N}$
           is a homomorphism. This requires defining the arithmetic
           relations and operations directly on the states
           and showing that they satisfy the relevant axioms.}
           \begin{equation}\label{equalA}
            |\g,0,h,s\rangle_{k,g} =_{A,k,g}|\g^{\p},0,h',s^{\p}
            \rangle_{k,g}\end{equation} holds if $\g^{\p}=\g$ and $s'=s$ up
            to leading and trailing $0s.$\footnote{That is, for all $j,$
            If $j$ is in  both $[l,u]$ and $[l',u'],$ then
            $s(j,h)=s'(j,h').$  If $j$ is in  $[l,u]$ and not in $[l',u'],$
            then $s(j,h)=0.$ If $j$ is in  $[l',u']$ and not in $[l,u],$
            then $s'(j,h')=0.$ The domains of $s$ and $s'$ are $[l,h;u,h]$ and
            $[l',h';u',h'].$}

            Arithmetic ordering $<_{A,k,g}$ on $N$, and on
         positive $I$ and $Ra$ states,
         \begin{equation}\label{deforderA}|+,0,h,s\rangle_{k,g}
         <_{A,k,g}|+,0,h',s^{\p}\rangle_{k,g}\end{equation} expresses the
         condition that the left hand state is arithmetically
         less than the right hand state.\footnote{The
         $<_{A,k}$ relation can be expressed by conditions on $s$
         and $s'$. Let $j_{max}$ and $j'_{max}$ be the largest
         $j$ values such that $s(j_{max},h)>0$ and $s'(j'_{max},h')>0$.
         Then $|+,0,h,s\rangle_{k,g} <_{A,k,g}|+,0,h',s^{\p}\rangle_{k,g}$
         if $j_{max} < j'_{max}$ or $j_{max}= j'_{max}$ and $s(j_{max,h}<
         s'(j'_{max},h')$.} The extension to zero and negative $I$ and
         $Ra$ states is given by the two conditions \begin{equation}
         \label{negordr} \begin{array}{l}|+,0,h,\bar{0}\rangle_{k,g,,}
         <_{A,k,g}|+,0,h',s'\rangle_{k,g}\mbox{ for all  $s'\neq\bar{0}$}\\
         |+,0,h,s\rangle_{k,g} <_{A,k,g}|+,0,h',s^{\p}\rangle_{k,g}\\
         \hspace{1cm} \rightarrow |-,0,h',s^{\p} \rangle_{k,g}<_{A,k,g}
         |-,0,h,s\rangle_{k,g}.\end{array}\end{equation} Here $\bar{0}$
         denotes a constant $0$ sequence.

         The $A$ subscript in  these relations emphasizes that
         these are arithmetic relations on the states. They are
         quite different from the usual quantum mechanical relations
         between states.  For instance, two states which differ by
         the number of leading or trailing $0s$ are arithmetically
         equal but are not quantum mechanically equal.

         The basic arithmetic operations on $Ra$ are $+,$ $-,$ $\times,$
         and a set of division operations, $\div_{\ell},$ one for
         each $\ell.$ This expresses the fact that the set of $k-ary$  rational
         string numbers is not closed under division when restricted to
         single finite length strings.  However it is closed under
         division to any finite accuracy, $k^{-\ell}.$
         For each $k,$ unitary operators for $+,$ $-,$ $\times,$ and
         $\div_{\ell}$ are represented by $\tilde{+}_{A,k,g},\;\tilde{-}_{A,k,g},$
         $\tilde{\times}_{A,k,g},$ and $\tilde{\div}_{A,k,g,\ell}.$
         These operators, acting on pairs of $q_{k}$ string states as
         input, generate an output triple consisting of the
         pair of input states and a result string state.

         To express this in a bit more detail, let $\tilde{O}_{A,k,g}$
         represent any of the four operation types, ($O=+,-,\times,\div_{\ell}.$)
         Then \begin{equation}\label{arithops}\begin{array}{l}
         \tilde{O}_{A,k,g}|\g,0,h, s\rangle_{k,g}|\gp,0,h',s'\rangle_{k,g}
         \\ \hspace{0.5cm}=|\g,0,h,s\rangle_{k,g}|\gp,0,h',s'
         \rangle_{k,g}|\g^{\p\p},0,h'', s^{\p\p}\rangle_{k,g,O_{A}}\end{array}
         \end{equation} The preservation of the input
         states is sufficient to ensure that the operators are
         unitary. The values of $h,h',h''$ are arbitrary except that
         they are all different.

         In these equations the states $|\gpp,0,h'',\spp\rangle_{k,g}$ with
         subscripts $O=+,-,\times,\div_{\ell}$ give the results of the arithmetic
         operations.  It is often useful to write them as
         \begin{equation}\label{addnplA}\begin{array}{l}
         |\g^{\p\p},0,h'',s^{\p\p}\rangle_{k,g,+}= |0,h'',(\gp,
         s'+_{A}\g,s)\rangle_{k,g},\\ |\g^{\p\p},0,h'',s^{\p\p}
         \rangle_{k,g,-}= |0,h'',(\gp,s' -_{A}\g, s)\rangle_{k,g},
         \\ |\g^{\p\p},0,h'',s^{\p\p}\rangle_{k,g,\times}=
         |0,h'',(\gp,s' \times_{A}\g,s)\rangle_{k,g}
         \\ |\g^{\p\p},0,h'',s^{\p\p}\rangle_{k,g,\div_{\ell}}=
         |0,h'',(\gp,s' \div_{A,\ell}\g,s)\rangle_{k,g}.\end{array}
         \end{equation} The subscript $A$ on these operations
         distinguishes them as arithmetic operations. They are different
         from the quantum operations of linear superposition, $+,-$
         and product, $\times$ with no subscripts.

         Extension of these operations to linear superposition
         states introduces entanglement.  Use of Eq. \ref{arithops}
         gives \begin{equation}\label{Opsipsip}\begin{array}{c}
         \tilde{O}_{A,k,g}\psi\psi'=\sum_{\g,h,s}\sum_{\g',h',s'}{_{k,g}\langle}
         \g,0,h,s|\psi\rangle_{k,g}\langle\g',0,h',s'|\psi'\rangle_{k,g}
         \\ \times |\g,0,h,s\rangle_{k,g} |\g',0,h',s'\rangle_{k,g}
         |\gpp,0,h'',s''\rangle_{k,g,O_{A}}.\end{array}\end{equation}

         Another operation that is essential for the axioms for $N$
         and is  useful for the others is that of the successor
         operation which corresponds to the $+1$ operation.  For
         $q_{k}$ string  states the definition can be expanded
         to include successor operators $\tilde{V}_{j}$ for each
         integer $j$.  The action of $\tilde{V}_{j}$ on a base $k$
         string state\begin{equation}\label{addj}\tilde{V}_{j}
         |\g,0,h,s\rangle_{k,g}=|\g',0,h,s'\rangle_{k,g}\end{equation}
         corresponds to the arithmetic addition of $k^{j}$ where $j$
         is any integer.  The usefulness of this operation is that the
         other arithmetic operations can be defined in terms of it.

         Also this definition provides an efficient way\footnote{The
         numbers of steps to implement the arithmetic operations are
         polynomial in the qukit string lengths.} to implement
         the arithmetic operations \cite{BenEIPSRN}.  This follows from
         the observations that for each $k$\begin{equation}\label{Njk}
         \tilde{V}_{j}^{k}=\tilde{V}_{j+1}\end{equation} and that the
         implementation of  each $\tilde{V}_{j}$ is efficient. Also
         implementation of the various arithmetic operations by use
         of the $\tilde{V}_{j}$ is efficient.

         \subsection{Transformations of Representations}\label{TR}

         As was noted earlier, the Fock space, $\mathcal{F}_{k}$, is
         spanned by the basis, $\mathcal{B}_{k,g},$ that is the set
         of all finite tuples of states of finite $q_{k}$ strings.
         Ultimately, $\mathcal{B}_{k,g}$ consists of sums and
         products of the individual $q_{k}$ bases,
         $\mathcal{B}_{k,h,j,g}$ where $\mathcal{B}_{k,h,j,g}$ is a
         set of $k$ single $q_{k}$ states that spans the $k$ dimensional
         Hilbert space $\mathcal{H}_{j,h}$ for site $j,h.$

         As is well known there are an infinite number of choices
         for a basis set in a Hilbert space. Here
         $\mathcal{B}_{k,h,j,g}$ denotes one choice. A choice of a
         basis set for each $\mathcal{H}_{j,h}$ is equivalent to a
         gauge fixing.  Thus a basis choice for each $j,h$ corresponds to a
         particular gauge choice at $j,h.$  The subscript $g$ represents
         a a gauge fixing function where for each integer pair $j,h,$
         \begin{equation}\label{gjhB}g(j,h)=\mathcal{B}_{k,j,h,g}.
         \end{equation}

         In what follows it is quite useful to treat
         $\mathcal{B}_{k,g}$ and $\mathcal{F}_{k}$ together. They
         will be denoted as $\mathcal{FB}_{k,g}.$ One reason for
         this is that the arithmetic relations and operations,
         which are needed to prove that the states $|\g,0,h,s\rangle_{k,g}$
         represent numbers, in $N$, $I$, and $Ra,$  are
         defined on the states in $\mathcal{B}_{k,g}$ and extend
         by linearity to states in $\mathcal{F}_{k}.$

         The  arguments given so far show that the set of all
         $\mathcal{FB}_{k,g}$ form a space parameterized
         by  a space of $2-tuples$ $(k,g).$ Here $k$ is a base and $g$
         is a gauge fixing function defined by Eq. \ref{gjhB}.
         Transformations $(k,g)\rightarrow (k',g')$ on
         the parameter space induce transformations
         $\mathcal{FB}_{k,g}\rightarrow \mathcal{FB}_{k',g'}$
         on the representation space. The two transformations of
         interest are the $k$ changing transformations
         $\tilde{W}_{k',k}$ and the gauge transformations $U_{k}.$
         Gauge transformations of the sign qubit are ignored here
         although they could be easily included.

         The gauge transformation, $U_{k}$ is a $U(k)=U(1)\times
         SU(k)$ valued function on $I\times I.$  $U_{k}$ is
         global if $U_{k}(j,h)$ is independent of $j,h.$  Otherwise
         it is local.   The action of $U_{k}$ changes the basis set
         or state reference frame for each qukit \cite{Enk} in that
         \begin{equation}\label{UBg}U_{k}(j,h)B_{k,j,h,g}=B_{k,j,h,g'}.
         \end{equation} holds for each $g$.

         One can use the definition of $U_{k}$ to define gauge transformation
         operators on $\mathcal{B}_{k,h,g}$ and $\mathcal{B}_{k,g}$.
         Here notation will be abused in that $U_{k}$ will represent
         all these transformations.  It will be clear from context
         which is meant.

         The action of $U_{k}$ on a state $|\g,0,h,s,\rangle_{k,g}$
         and the individual A-C operators is given by \begin{equation}
         \label{Urast}\begin{array}{l}
          |\g,0,h,s\rangle_{k,g'}=U_{k}|\g,0,h,s\rangle_{k,g}\\ \hspace{0.5cm} =
          \cd_{\g,0,h} U_{k}(u,h)(\ad_{k})_{s(u),u,h}
          \cdots U_{k}(l,h)(\ad_{k})_{s(l),l,h}|0\rangle \\ \hspace{1cm}=
          \cd_{\g,0,h}((\ad_{k})_{U_{k}(u,h)})_{s(u),u,h}\cdots((\ad_{k})_
          {U_{k}(l,h)})_{s(l),l,h}|0\rangle\end{array}\end{equation}
          where\begin{equation}\label{adU}\begin{array}{c}
          ((\ad_{k})_{U_{k}(j,h)})_{\alpha,j,h}=U_{k}(j,h)(\ad_{k})_{\alpha,j,h}=
          \sum_{\beta}U_{k}(j,h)_{\alpha,\beta}(\ad_{k})_{\beta,j,h}\\
          ((a_{k})_{U_{k}(j,h)})_{\beta,j,h}=(a_{k})_{\beta,j,h}U^{\dag}_{k}(j,h)=
          \sum_{\alpha}U^{*}_{k}(j,h)_{\alpha,\beta}a_{\alpha,j,h}
          \end{array}\end{equation} These results are based on the
          representation of $U_{k}(j,h)$ as \begin{equation}\label{Uexpand}
          U_{k}(j,h)=\sum_{\alpha,\beta}(U_{k}(j,h))_{\alpha,\beta}
          (\ad_{k})_{\alpha,j,h}(a_{k})_{\beta,j,h}.\end{equation} Here
          $((\ad_{k})_{U_{k}(j,h)})_{\alpha,j,h}$ is the creation
          operator for  $q_{k}$ in the state $|\alpha,j,h\rangle_{k,g'}$
          in the basis $B_{k,j,h,g'}$ just as $(\ad_{k})_{\alpha,j,h}$
          is the creation operator for $q_{k}$ in the state
          $|\alpha,j,h\rangle_{k,g}$ in the basis $B_{k,j,h,g}.$

         The base changing operator $\tilde{W}_{k',k}$ is more
         complex. If $\tilde{W}_{k',k}$ is defined on the state
         $|\g,0,h,s\rangle_{k,g}$, then \begin{equation}\label{Wssp}
         |\g,0,h,s'\rangle_{k',g}=\tilde{W}_{k',k}|\g,0,h,s\rangle_{k,g}
         \end{equation} represents the same number in the base $k'$
         representation as $|\g,0,h,s\rangle_{k}$ does in the base
         $k$ representation. This a nontrivial requirement because
         one needs to specify what is meant by "the same number as".
         In particular it means that all number theoretic properties
         are valid for $|\g,0,h,s\rangle_{k},$ if and only if they
         are valid for $|\g,0,h,s'\rangle_{k'}.$

         For any $k',k,$ the operator $\tilde{W}_{k',k}$ is defined
         for all natural number and integer qukit string states. For qukit
         string states that represent rational numbers the domain
         and range of $\tilde{W}_{k',k}$ depend on the relations
         between the prime factors of $k$ and $k'.$  The domains
         and ranges for the different cases are summarized by the
         following relations \cite{BenSQTRN}. Let $PF(k)$ denote
         the prime factors of $k$. Then\begin{equation}\label{domranW}
         \begin{array}{l}\mbox{If $PF(k)\bigcap PF(k')=0$ then the
         domain and range of $\tilde{W}_{k',k}$} \\ \hspace{0.5cm}
         \mbox{are the integer subspaces of $\mathcal{FB}_{k,g}$ and
         $\mathcal{FB}_{k',g}$} \\ \mbox{If $PF(k)\subset PF(k')$ then
         $\tilde{W}_{k',k}\mathcal{FB}_{k,g}\subset
         \mathcal{FB}_{k',g},$}\\ \mbox{If $PF(k)\supset PF(k')$ then
         $\tilde{W}_{k',k}\subset\mathcal{FB}_{k,g}=
         \mathcal{FB}_{k',g},$}\\ \mbox{If $PF(k),PF(k')$ each have elements
         not in the other and}\\ \mbox{}\hspace{0.8cm}\mbox{share
         elements in common, then } \tilde{W}_{k',k}\subset\mathcal{FB}_{k,g}=
         \subset\mathcal{FB}_{k',g},\\ \mbox{If $PF(k)= PF(k')$ then
         $\tilde{W}_{k',k}\mathcal{FB}_{k,g}=\mathcal{FB}_{k',g}.$}
         \end{array}\end{equation}In  the above $\subset\mathcal{FB}_{k,g}$
         denotes a subspace of $\mathcal{FB}_{k,g}$ that contains the
         integer representations.  In all these cases, if the state
         $|\g,0,h,s\rangle_{k,g}$ is in the domain of $\tilde{W}_{k',k},$
         then the base $k'$ state, $\tilde{W}_{k',k}|\g,0,h,s
         \rangle_{k,g},$ represents the same rational number as does
         $|\g,0,h,s\rangle_{k,g}.$

         The case where $PF(k)=PF(k')$ is of special interest
         because for each $k$ there is a smallest $k'$ that has
         the same prime factors as $k$.  If\begin{equation}\label{kprfa}
         k=p_{j_{1}}^{h_{1}}\cdots p_{j_{n}}^{h_{n}},\end{equation}
         then the smallest $k'$ is given by \begin{equation}\label{kpprfa}
         k'=p_{j_{1}}\cdots p_{j_{n}}. \end{equation} Here
         $p_{j_{a}}$ for $a=1,2,\cdots,n$ is the $j_{a}th$ prime
         number. This shows that for each finite subset $S$ of
         primes, there is a set $[k_{S}]$ of bases such that for any
         pair $k,k' \epsilon [k_{S}],$ $\tilde{W}_{k',k}$ is defined
         everywhere on $\mathcal{FB}_{k,g}$ and
         $\tilde{W}_{k',k}\mathcal{FB}_{k,g}=\mathcal{FB}_{k',g}.$

         A special case of this consists of the values $k_{n}$
         whose factors are the first $n$ primes, each to the first
         power, \begin{equation}\label{kbase} k_{n}=p_{1}p_{2}\cdots
         p_{n}=2\times 3\times\cdots\times p_{n}.\end{equation} The
         sets $[k_{n}]$ are of interest here because, if $n<m,$ then
         $[k_{n}]\subset [k_{m}].$ The limit properties, as
         $n\rightarrow \infty,$ of $[k_{n}]$ and $\tilde{W}_{k',k_{n}}$
         are open for investigation.

         It should also be noted that the definitions of both
         $U_{k}$ and $\tilde{W}_{k',k}$ extend by linearity to
         linear superpositions of qukit string states.  If $\psi
         =\sum_{\g,h,s}c_{\g,h,s}|\g,0,h,s\rangle_{k,g},$ then
         \begin{equation}\label{linsup}\begin{array}{l} U_{k}\psi
         =\sum_{\g,h,s}c_{\g,h,s}U_{k}|\g,0,h,s\rangle_{k,g} \\
         \tilde{W}_{k',k}\psi =\sum_{\g,h,s}c_{\g,h,s}\tilde{W}_{k',k}
         |\g,0,h,s\rangle_{k,g}.\end{array}\end{equation} The validity
         of the second equation is restricted to the case where all
         component states with nonzero coefficients are in the domain of
         $\tilde{W}_{k',k}.$

         It is of interest to note that there is in general no
         commutation relation between $U_{k}$ and
         $\tilde{W}_{k,k'}.$ The one exception is the case when
         $k'=k^{n}$ for some $n$. However, for each pair $k,k'$ for
         which $\tilde{W}_{k,k'}$ is defined everywhere on
         $\mathcal{F}_{k},$ and for each pair $U_{k},U'_{k'}$ one can
         define a transformed operator\begin{equation}\label{UWUd}
         (\tilde{W}_{U',U})_{k',k}=U'_{k'}\tilde{W}_{k',k}U^{\dag}_{k}.
         \end{equation}

         This operator takes a transformed state
         $U_{k}|\g,h,s\rangle_{k,g}$ to a base $k'$ state
         $$(\tilde{W}_{U',U})_{k',k}U_{k}|\g,h,s\rangle_{k,g}=
         U'_{k'}\tilde{W}_{k'.k}|\g,h,s\rangle_{k,g}$$ that
         represents the same number in the $k',g'$ representation as
         $|\g,h,s\rangle_{k,g}$ does in the $k,g$ representation.
         The steps in the representation transformations are
         \begin{equation}\label{UtoWtoUp}(k,g)\stackrel{U_{k}}
         {\rightarrow}(k,g_{1})\stackrel{\tilde{W}_{k,k'}}{\rightarrow}
         (k',g_{1})\stackrel{U'_{k'}}{\rightarrow}(k',g').\end{equation}
         Note that basis or gauge choice $g_{1}$ chosen for the base
         $k$ states is used to label the gauge choice for base $k'$
         states that are connected by $\tilde{W}_{k,k'}.$

         \subsection{Transformations of Arithmetic Relations
         and Operations}\label{TARO} The arithmetic relations and
         operations transform in the expected way under the action
         of $\tilde{W}_{k,k'}$ and $U_{k}.$ One has\begin{equation}
         \label{ReWW}\begin{array}{c} =_{A,k',g}=(\tilde{W}_{k,k'}
         =_{A,k,g}\tilde{W}^{\dag}_{k,k'})\\\leq_{A,k',g}=\tilde{W}_{k,k'}
         \leq_{A,k,g}\tilde{W}^{\dag}_{k,k'}\end{array}\end{equation}
         for the relations and\begin{equation}\label{OpWW}
         \begin{array}{l}O_{A,k',g}=\tilde{W}_{k,k'}\times
         \tilde{W}_{k,k'}\times\tilde{W}_{k,k'}\\\hspace{1cm}\times
         O_{A,k,g}\tilde{W}^{\dag}_{k,k'}\times\tilde{W}^{\dag}_{k,k'}
         \end{array}\end{equation} for $O= +,\times, -,\div_{\ell}.$

         These transformations of relations and operations hold
         without restrictions if and only if If $k$ and $k'$ have
         the same prime factors. If this is not the case, then the
         restrictions expressed by Eq. \ref{domranW} apply here.
         In the worst case where $k$ and $k'$ are relatively prime,
         the transformations are restricted to the integer
         subspaces of $\mathcal{F}_{k}$ and $\mathcal{F}_{k'}.$ The
         presence of three transformation operators on the left of
         $O_{A,k,g}$ and two to the right accounts for the fact that
         $O_{A,k,g}$ preserves the two input strings and creates a
         third.

         One has similar relations for the gauge transformations of
         relations and operations.\begin{equation}\label{ReUU}
         \begin{array}{c} =_{A,k,g'}=(U_{k}=_{A,k,g}U^{\dag}_{k})
         \\\leq_{A,k',g}=U_{k}\leq_{A,k,g}U^{\dag}_{k}\end{array}
         \end{equation} for the relations and\begin{equation}\label{OpUU}
         \begin{array}{l}O_{A,k,g'}=U_{k}\times U_{k}\times U_{k}\\
         \hspace{1cm}\times O_{A,k,g}U^{\dag}_{k}\times U^{\dag}_{k}
         \end{array}\end{equation} for $O= +,\times, -,\div_{\ell}.$

         \subsection{Unary Representations}\label{k1}
         So far all number bases have been considered except one,
         the value $k=1.$ The $k=1$ string representations are
         called unary representations.  These are not usually
         considered, because basic arithmetic operations on these numbers
         are exponentially hard. For instance the number of steps
         needed to add two unary numbers is proportional to the
         values of the numbers and not to the logarithms of the
         values. However, even though they are not used
         arithmetically, they are always present in an interesting way.

         To see this one notes that $k=1$ representations are the
         only ones that are  extensive, all others are
         representational. The representational property for
         $k\geq 2$ base states of a qukit string means that a number
         represented by a state has nothing to do with the properties
         of the string state. The number represented by the
         state, $|672\rangle,$  of a string of $3$ $q_{10}'s$ is
         unrelated to the properties of the qukits in  the state.

         The extensiveness  of a unary representation means that
         any collection of systems is an unary representation of a
         number that is the number of systems in the collection.
         There are many examples. A system of spins on a lattice
         is an unary representation of a number, that is the number
         of spins in the system. A gas of particles in a box is an unary
         representation of a number, that is the number of particles
         in the box. The qukit strings that play such an important role in this
         paper are unary representations of numbers, that are the
         number of qukits in the strings. A single qukit is an unary
         representation of the number $1$.

         The omnipresence of unary representations relates to
         another observation that $1$ is the only number that is a
         common factor of all prime numbers and of all numbers. So
         it is present as a factor of any base.  This ties in with
         the fact that unary representations of numbers are possible
         only for natural numbers and integers.\footnote{Non integer
         rational numbers require pairs of unary representations.
         However, pairs are not being considered here.} Also there is
         the related observation that, for any pair $k,k',$ the domain
         and range of $\tilde{W}_{k',k}$ include the integer subspaces
         of $\mathcal{F}_{k}$ and $\mathcal{F}_{k'},$ and if $k,k'$
         have no prime factors in common, $\mathcal{F}_{k}$ and
         $\mathcal{F}_{k'}$ are the domain and range of $\tilde{W}_{k',k}.$

         The extensiveness of unary representations supports
         the inclusion of the $U(1)$ factor in the definition of  $U_{k}$
         as a $U(1)\times SU(k)$ valued function on $I\times I.$ As
         a very simple example, a state $(\ad_{k})_{\alpha,(i,j)}
         |0\rangle$ of a qukit at location $(i,j)$ is an
         unary representation of the number $1$. Multiplication of
         this state by a phase factor $e^{i\theta_{i,j}}$ is a
         transformation that gives another state that is also an
         unary representation of the number $1$.

         This argument extends to states of strings of qukits. A
         phase factor associated with any state of a string of
         $q_{k}$ at sites $(l,h),\cdots (u,h)$ is a product of the
         phase factors associated with each of the $q_{k}$ in the
         string.  If $e^{i\theta_{j,h}}$ is a phase factor for the
         state of the $q_{k}$ at site $(j,h),$ then $ e^{i\Theta_{
         [(l,h),(u,h)]}},$ where $\Theta_{[(l,h),(u,h)]}=\sum_{j=l}^{u}
         \theta_{j,h},$ is the phase factor for the state of the
         $q_{k}$ string in the site interval $[(l,h),(u,h)]$.

         As is well known, multiplying any state by a phase factor
         gives the same state as far as any physical meaning is
         concerned.  However here one can have linear superpositions
         of states of strings of $q_{k}$ both at different locations
         and of different length strings.  In these cases the phase
         factors do matter to the extent that they  can change the
         relative phase between the components in the superposition.

          \subsection{Composite and Elementary Qukits}\label{CEQ}

           So far the qukit components of strings are considered to
           be different systems for each value of $k.$ A $k$ qukit is
           different from a $k^{\p}$ qukit just as a spin $k$ system
           is different from a spin $k^{\p}$ system. This leads to a
           large number of different qukit types, one for each value
           of $k.$ However, the dependence of the properties of the
           base changing operator $\tilde{W}_{k',k}$ on the prime
           factors of $k$ and $k'$ suggests that instead one consider qukits
           $q_{k}$ as composites $q_{c_{k}}$ of prime factor qukits
           $q_{p_{n}}$. In general the relation between the
           base $k$ $q_{k}$ and the composite base $c(k)$
           $q_{c_{k}}$ is given by\begin{equation}\label{qck}
           q_{c_{k}}=q_{p_{j_{1}}}^{h_{1}}q_{p_{j_{2}}}^{h_{2}}\cdots
           q_{p_{j_{n}}}^{h_{n}}.\end{equation} where (Eq. \ref{kprfa})
           $$k=p^{h_{1}}_{j_{1}}p^{h_{2}}_{j_{2}}\cdots p^{h_{n}}_{j_{n}}$$.
           Simple examples of this for $k=10$ and $18$ are
           $q_{c_{10}}=q_{2}q_{5}$ and $q_{c_{18}}=q_{2}q_{3}q_{3}.$

           The observation that for each $k$ there is a smallest $k'$
           with the same prime factors and its relevance to the
           properties of $\tilde{W}_{k',k}$ suggest the importance
           of the $q_{c_{k'}}$ where the powers of the prime factors
           are all equal to $1$(Eq. \ref{kpprfa}) \begin{equation}
           \label{qckp} q_{c_{k'}}=q_{p_{j_{1}}}q_{p_{j_{2}}}\cdots
           q_{p_{j_{n}}}.\end{equation} A particular example of this
           for $k_{n},$ the product of the first $n$ prime numbers, is
           shown by (Eq. \ref{kbase})\begin{equation} \label{qckn}
           q_{_{c_{k_{n}}}}=q_{2}q_{3}q_{5}\cdots q_{p_{n}}.
           \end{equation}

           These considerations suggest a change of emphasis in
           that one should regard  prime number qukits $q_{p_{n}}$
           as basic or elementary and the qukits $q_{k}$ as composites
           of the elementary ones. In this case one would want to consider
           possible physical properties of the elementary qukits and
           how they interact and couple together to form composites.
           This is a subject for future work.  It
           is, however, intriguing to note that if the prime number
           $q_{p_{n}}$ are considered as spin systems with spin
           $s_{n}$ given by $2s_{n}+1=p_{n}$, then there is
           just one fermion, $q_{2}$.  All the others are bosons.

           As was the case for strings of $q_{k},$ one wants to
           represent numbers by states of finite strings of composite
           $q_{c_{k}}$. In general, this involves replacing the $k$
           dimensional Hilbert space $\mathcal{H}_{k}$ at each site
           in $I\times I$ by a product space\begin{equation}\label{Hck}
          \mathcal{H}_{c_{k}}=\mathcal{H}_{p_{j_{1}}}^{h_{1}}\otimes
          \cdots\otimes\mathcal{H}_{p_{j_{n}}}^{h_{n}},\end{equation}and
          then following the development in the previous sections to
          describe number states. In particular the gauge fixing would
          apply to each component space in Eq. \ref{Hck} for each
          location in $I\times I.$

          The requirement that states of the form
          $|\g,0,h,s'\rangle_{c_{k'},g}$ represent numbers is
          based on an ordering of the basis states of $q_{c_{k}}$,
          or, what is equivalent, an ordering of the  $n-tuples$ in
          the range set of $s'$.   The definitions of
          arithmetic relations and operations for these states must
          respect the ordering and they must satisfy the relevant axioms
          and theorems for the type of number being considered.

          The description of the transformation operations
          $\tilde{W}_{k',k}$ and $U_{k'}$ can be extended to apply to
          the composite qukit strings. The base changing operator
          $\tilde{W}_{c_{k'},c_{k}}$ changes states of $q_{c_{k}}$
          strings to states of $q_{c_{k'}}$ strings that should
          represent the same number.  Note that the expression of
          $\tilde{W}_{c_{k'},c_{k}}$ in terms of sums of products of
          AC operators will include the annihilation of many
          component elementary qukits in $q_{c_{k}}$ and creation
          of many that are components of $q_{c_{k'}}.$

          The description of gauge transformations $U_{c_{k}}$
          applied to states of $q_{c_{k}}$ is interesting. If
          $q_{c_{k}}$ is composed of elementary $q_{p_{j}}$ as given
          by Eq. \ref{qckp}, then $U_{c_{k}}$ is a map from $I\times
          I$ to elements of $U(p_{j_{1}})\times\cdots\times
          U(p_{j_{n}}).$ Here $U(p_{j_{i}})$ is the unitary group
          of prime dimension $p_{j_{i}}.$ For the special case of
          Eq. \ref{qckn}, $U_{c_{k_{n}}}$ takes values in
          $U(p_{1})\times\cdots\times U(p_{n}).$ respectively. Since
          $U(p_{j}) =U(1)\times SU(p_{j})$ the values of
          $U_{c_{k_{n}}}$ can be represented as elements of
          \begin{equation}\label{SUprod}\begin{array}{l}
          U(1)\times SU(p_{1})\times SU(p_{2})\times\cdots\times
          SU(p_{n})\\ \mbox{}\hspace{1cm}=U(1)\times SU(2)\times SU(3)\times
          SU(5)\times\cdots\times SU(p_{n}).\end{array}\end{equation} Here the
          phase factor elements in $U(1)$ for each elementary qukit
          have been combined into one phase factor for the composite
          $q_{c_{k_{n}}}.$

          This brief description of composite and elementary qukits
          shows that this may be an interesting approach to examine
          further. Problems to investigate include the nature of the
          coupling of elementary qukits to form a composite,
          invariance of properties of composite qukit string states
          under the action of  $U_{k}$, particularly of
          $U_{c(k_{n})}$, and other aspects.

          The discussion so far suggests that, as far as quantum
          theory representations of natural numbers, integers, and
          rational numbers are concerned, it is sufficient to limit
          components of gauge transformations to  products of elements
          of $U(1)$ and products of elements of $SU(p)$ groups
          where $p$ is a prime number. Furthermore it is sufficient
          that, for each prime $p,$ elements of $SU(p)$ occur at most
          once in the product. It is also sufficient to limit
          components to products of the form of Eq. \ref{SUprod} for
          $n=1,2,\cdots$ as these will include representations for
          all rational numbers.

         \section{Quantum Representations of Real Numbers}\label{QRRN}

         Here quantum representations of real numbers are described
         as equivalence classes of  sequences of  base $k\geq 2$
         qukit ($q_{k}$) string states  that satisfy the Cauchy
         condition.\footnote{This extends earlier work \cite{BenRRCNQT}
         on real number representations that was limited to $k=2$.}
         Sequences of states are defined to be functions $\Psi$
         from the natural numbers to states in $\mathcal{F}_{k}.$ If
         the states in the range set of $\Psi$ are all basis states
         in $\mathcal{B}_{k,g}$, then $\Psi(n)=|\g_{n},h_{n},s_{n}
         \rangle_{k,g}$.   The values of $h_{n}$ in the states
         $|\g_{n},h_{n},s_{n} \rangle_{k}$ are all different
         in that $m\neq n\rightarrow h_{m}\neq h_{n}.$ This is
         needed because one must be able to distinguish
         $\Psi(n)$ from $\Psi(m).$ Here and from now on  the
         location $0$ of the sign qubit in $|\g,0,h,s\rangle_{k,g}$
         is suppressed as it is always the same.

         These sequences extend classical representations in that
         the $\mathcal{B}_{k,g}$ valued sequences correspond to
         classical states of kit sequences. However sequences of
         linear superposition states have no classical correspondences.

         \subsection{The Cauchy Condition for State Sequences}

         The definition of the Cauchy condition for sequences
         of $q_{k}$ string states is a translation into
         quantum mechanics of a definition in mathematical
         analysis textbooks \cite{Burkhill}.  To this end
         let $\Psi$ be a $\mathcal{B}_{k,g}$ valued
         sequence of  $q_{k}$ string states. The sequence
         $\Psi$ satisfies the Cauchy
         condition if \begin{equation}\label{cauchy}\begin{array}{l}
         \hspace{0.5cm}\mbox{ For each $\ell$ there is a $p$
         where for all $j,m>p$} \\ \hspace{1cm}
         |(|\Psi(j)-_{A,k,g}\Psi(m)|_{A,k,g})\rangle_{k,g}
         <_{A,k,g}|+,-\ell\rangle_{k,g}.\end{array} \end{equation}

         Here $|(|\Psi(j)-_{A,k,g}\Psi(m)|_{A,k,g})
         \rangle_{k,g}$ is the basis state that is the base $k$ arithmetic
         absolute value of the state resulting from the arithmetic
         subtraction of $\Psi(m)$ from $\Psi(j).$ The Cauchy condition
         says that this state is arithmetically less than or equal to the
         state $|+,-\ell\rangle_{k,g}=|+,0_{[0,-\ell+1]}1_{-\ell}\rangle_{k,g}$
         for all $j,m$ greater than some $p$. Here $|+,-\ell\rangle$
         represents the number $k^{-\ell}.$ The
         subscripts $A,k,g$ are used to indicate that the operations
         are arithmetic and are defined for base $k$ string states in
         $\mathcal{B}_{k,g}.$ They are not the usual quantum theory
         operations.\footnote{When it is desired to emphasize the
         dependence of the definition of the Cauchy condition on
         $g,$ Eq. \ref{cauchy} will be referred to as the g-Cauchy
         condition.}

         The Cauchy condition can be extended to sequences  of linear
         superpositions of $q_{k}$ string states. Let $\Psi(n)
          =\sum_{\g,h,s}|\g,h, s\rangle_{k,g}\langle\g ,h,s|\Psi(n)\rangle.$
          The probability that the arithmetic absolute value
          of the arithmetic difference between $\Psi(j)$ and
          $\Psi(m)$ is arithmetically less than or equal to
          $|+,-\ell\rangle$ is given by\begin{equation}\label{Pjml}
          \begin{array}{l}P_{j,m,\ell}= \sum_{\g,h,s}
          \sum_{\gp,h',s'} |\langle\g,h,s|\Psi(j)\rangle
          \langle\gp,h',s'|\Psi(m)\rangle|^{2}: \\ \hspace{2cm}
          |(|\g,h,s-_{A,k,g}\gp,h' ,s'|_{A,k,g})\rangle_{k,g} \leq_{A,k,g}
          |+,-\ell\rangle_{k,g}. \end{array}\end{equation} The sum is
          over all $|\g,h,s\rangle,|\gp,h',s'\rangle$ that satisfy the
          statement in the second line of the above equation.

          The definition of the probability $P^{\Psi}$ that $\Psi$ satisfies
          the Cauchy condition is obtained from the values of $P^{\Psi}_{n,m,\ell}$
          by taking account of the quantifiers in the definition in Eq.
          \ref{cauchy}.  To this end define the
          probabilities $P^{\Psi}_{p,\ell},$  $P^{\Psi}_{\ell},$ and
          $P^{\Psi}$ by\begin{equation}\label{PpsiC}\begin{array}{c}
          P^{\Psi}_{p,\ell}=\inf_{n,m>p}P^{\Psi}_{n,m,
          \ell}\\P^{\Psi}_{\ell}=\limsup_{p\rightarrow\infty}P^{\Psi}_{p,\ell}
          =\lim_{p\rightarrow\infty}P^{\Psi}_{p,\ell}\\P^{\Psi}_{C}=
          \liminf_{\ell\rightarrow\infty}P^{\Psi}_{\ell}=\lim_{\ell
          \rightarrow\infty}P^{\Psi}_{\ell}.\end{array}\end{equation}

          This definition is based on the structure
          of the Cauchy condition in Eq. \ref{cauchy}. It shows
          that the asymptotic values of $P^{\Psi}_{n,m,\ell}$ as
          $m,n\rightarrow\infty$ are important. The values for
          any particular $m,n$ or finite set $\{m,n\}$ of values
          (with $\ell$ fixed) for each $\ell$ are not important.
          The structure also shows that $P^{\Psi}_{p,\ell}$ is a
          non decreasing function of $p$ and that $P^{\Psi}_{\ell}$
          is a non increasing function of $\ell.$

          The sequence $\Psi$ is said to
          be a Cauchy sequence  if  $P_{\Psi}$ is equal
          to $1$. A necessary and sufficient condition for this to occur is that
          $P^{\Psi}_{n,m,\ell}\rightarrow 1$ as $n,m\rightarrow\infty$
          for each $\ell.$ That is,\begin{theorem}
          $P_{\Psi}=1\Leftrightarrow\lim_{m,n\rightarrow\infty}
          P^{\Psi}_{n,m,\ell}=1 \mbox{ for each
          $\ell.$}$ \end{theorem} \underline{Proof}
          \emph{sufficiency}: Obvious as probabilities are bounded
          above by $1$, One has $P^{\Psi}_{p,\ell}=1$ for each $p$
          and $\ell$.  This gives $P^{\Psi}_{\ell} =1$ for each
          $\ell.$\\ \\\emph{necessity}: Assume $P_{\Psi}=q<1.$
          From the definition of $P^{\Psi}_{\ell}$ one sees that it
          approaches $q$ from above as $\ell$ increases. It follows
          that for sufficiently large $\ell$, $P^{\Psi}_{p,\ell}$,
          which is non decreasing is bounded from above by $q$ as
          $p$ increases. It follows from the definition of
          $P^{\Psi}_{p,\ell}$ that either
          $\lim_{m,n\rightarrow\infty}P^{\Psi}_{n,m,\ell}$ does not
          exist or it exists and is $\leq q.$ QED

          There are many examples of sequences  $\Psi$ that are
          Cauchy with probability $1.$ A simple example is the
          following: Let $s$ be a $0,1,\cdots,k-1$ valued function
          on the non positive integers $[0,-\infty].$  Define $\Psi(n)$ by
          \begin{equation}\label{Psiex1}\Psi(n)=|+,0,h,s_{[0,-n+1]}
          \rangle_{k}\times\frac{1}{\sqrt{k}}\sum_{j=0}^{k-1}|-n,h,j
          \rangle_{k}.\end{equation} Here $|j,h,-n\rangle_{k}$
          denotes a $q_{k}$ at site $-n,h$ in state $j$.

          The sequence $\Psi$ is Cauchy with probability $1$
          because $P^{\Psi}_{n,m,\ell} =1$ for all $m,n>\ell.$
          Also this example does not correspond to any classical
          Cauchy sequence of rational numbers. Additional examples
          are given in \cite{BenRRCNQT}.

          \subsection{Basic Relations on State
          Sequences}\label{BRSS}
          One way to proceed is to define the field relations and
          operations on equivalence classes of sequences and show
          that these satisfy the real number axioms. However this
          method does not make clear the relation between the basic
          arithmetic relations on the $q_{k}$ string states and
          those on the equivalence classes.  The method used here is
          to define the basic relations and operations on the
          sequences in terms of the relations and operations on the
          $q_{k}$ string states and use them to define the
          equivalence classes and field relations and operations on
          the classes.

          As a piece of nomenclature let $Re_{i,S,k,g}$ denote the
          two relations  where $Re_{i,S,k,g}$ is $=_{S,k,g}$ (equality)
          for $i=1$ and $Re_{i,S,k,g}$ is $ <_{S,k,g}$ (less than)
          for $i=2.$ The simplest definition one thinks of is an
          elementwise definition:\begin{equation}\label{1stReiS}
          \Psi(n) Re_{i,S,k,g}\Psi'\Leftrightarrow \forall{n}\Psi
          Re_{i,A,k,g}\Psi'(n). \end{equation} Here $Re_{i,A,k,g}$
          corresponds to the two relations on finite $q_{k}$ string
          states.

          These definitions are unsatisfactory in that they are too strong.
          For $i=1$ this definition gives the result that for most
          $\mathcal{F}_{k}$ valued sequences, $\Psi,$ the probability
          that $\Psi=_{S,k,g}\Psi$ is $0.$ This holds even if $\Psi$
          is Cauchy. For $i=2$ the definition of $<_{S,k,g}$ does
          not have the right asymptotic properties.

          A better definition of $=_{S,k,g}$ is an asymptotic
          definition. Let $\Psi$ and $\Psi'$ be $\mathcal{B}_{k,g}$
          valued sequences. Then\begin{equation}\label{EqAS}
          \begin{array}{l}\Psi =_{\infty,k,g}\Psi'\Leftrightarrow \\\hspace{1cm}
          \forall{\ell}\exists{p}\forall{j,m>p}|(|\Psi(j)-_{A,k,g}
          \Psi'(m)|_{A,k,g})\rangle_{k} <_{A,k,g}|+,-\ell\rangle_{k,g}.
          \end{array}\end{equation} This definition is the same as
          the Cauchy condition of Eq. \ref{cauchy} except that
          $\Psi'(m)$ replaces $\Psi(m).$ This definition says
          nothing about whether specific elements of $\Psi$ equal
          the corresponding ones of $\Psi'.$ It says that the
          elements of $\Psi$ and $\Psi'$ must approach each other
          asymptotically. It is easy to show that this definition
          satisfies the requirement for a definition of equality.
          It is reflexive, symmetric, and transitive.

          An asymptotic definition of ordering is given by
          \begin{equation}\label{LtAS}\begin{array}{l}\Psi <_{\infty,k,g}
          \Psi'\Leftrightarrow \\\hspace{1cm}\exists{\ell}\exists{p}
          \forall{j,m>p}|\Psi'(j)-_{A,k,g}\Psi(m)\rangle_{k}
          >_{A,k,g}|+,-\ell\rangle_{k}.\end{array}\end{equation} This
          is also an asymptotic definition in that it says that
          $\Psi$ is less than $\Psi'$ if $\Psi$ is asymptotically
          arithmetically less than $\Psi'$ by some fixed amount,
          $|+,-\ell\rangle_{k}.$ This definition differs from Eq.
          \ref{EqAS} in that $\forall{\ell}$ is replaced by
          $\exists{\ell}$, there is no arithmetic absolute value,
          and $<_{A,k,g}$ is replaced by $>_{A,k,g}.$

           These definitions can be extended to $\mathcal{F}_{j,k}$
           valued sequences. Let $\Psi$ and $\Psi'$ be sequences of
           this type.  Define $P_{n,m,\ell}^{\Psi =_{\infty}\Psi'}$ by,
           \begin{equation}\label{Ppsinml}\begin{array}{l}P^{\Psi
           =_{\infty}\Psi'}_{n,m,\ell}= \sum_{\g,h,s}\sum_{\g',h',s'} |d^{n}_{\g,h,s}
           f^{m}_{\g',h',s'}|^{2}: \\\hspace{1cm}|(|(\g,h,s)-_{A,k,g}
           (\g',h',s')|_{A,k,g})\rangle_{k}\leq_{A,k,g}
          |+,-\ell\rangle_{k} \end{array}\end{equation} where
          \begin{equation}\label{dfpsi}\begin{array}{c}d^{n}_{\g,h,s}=
          \langle\g,h,s|\Psi(n)\rangle\\ f^{m}_{\g',h',s',}=\langle
          \g',h',s|\Psi'(m)\rangle.\end{array}\end{equation} Here
          $P_{n,m,\ell}^{\Psi =_{\infty}\Psi'}$ is the probability that
          $\Psi(n)$ and $\Psi'(m)$ satisfy the relation in the second
          line of Eq. \ref{Ppsinml}.

          Let $P^{\Psi=_{\infty}\Psi'}$ be the probability that
          $\Psi =_{\infty}\Psi',$ i. e. that  $\Psi$ equals $\Psi'$ $S$
          asymptotically. Here $P^{\Psi=_{\infty}\Psi'}$ is given by
          \begin{equation} \label{Ppsieq}\begin{array}{c} P^{\Psi
          =_{\infty}\Psi'}_{p,\ell}=\inf_{n,m>p} P^{\Psi=_{\infty}\Psi'}_{n,m,
          \ell}\\P^{\Psi=_{\infty}\Psi'}_{\ell}=\limsup_{p\rightarrow\infty}
          P^{\Psi=_{\infty}\Psi'}_{p,\ell} =\lim_{p\rightarrow\infty}
          P^{\Psi=_{\infty}\Psi'}_{p,\ell}\\P^{\Psi=_{\infty}\Psi'}=\liminf_{
          \ell\rightarrow\infty}P^{\Psi=_{\infty}\Psi'}_{\ell}=\lim_{\ell
          \rightarrow\infty}P^{\Psi=_{\infty}\Psi'}_{\ell}.\end{array}
          \end{equation}These equations are similar to those in Eq. \ref{PpsiC}
          because the quantifier setup in Eq. \ref{EqAS} is the same
          as that for the Cauchy condition in Eq. \ref{cauchy}. As was
          the case before, $P^{\Psi=\Psi'}_{p,\ell}$ is a non decreasing
          function of $p$ for each $\ell$  and $P^{\Psi=\Psi'}_{\ell}$ is
          a non increasing function of $\ell.$

          A similar result holds for the probability $P^{\Psi<_{\infty}
          \Psi'}$ that $\Psi$ is asymptotically less than $\Psi'.$  Eqs.
          \ref{LtAS} and \ref{dfpsi} give \begin{equation}\label{Ppsinml2}
          \begin{array}{l}P^{\Psi <_{\infty}\Psi'}_{n,m,\ell}= \sum_{\g,h,s}
          \sum_{\g',h',s'} |d^{n}_{\g,h,s}f^{m}_{\g',h',s'}|^{2}: \\ \hspace{1cm}
          |(\g',h',s')-_{A,k,g}(\g,h,s)\rangle_{k,g} \geq_{A,k,g}
          |+,-\ell\rangle_{k,g}. \end{array}\end{equation}

            From Eq. \ref{LtAS} one obtains results for
          $P^{\Psi<_{\infty} \Psi'}$ that are different from
          Eq. \ref{Ppsieq}:\begin{equation}
          \label{Ppsieq1}\begin{array}{c} P^{\Psi<_{\infty}\Psi'}_{p,\ell}=
          \inf_{n,m>p} P^{\Psi<_{\infty}\Psi'}_{n,m,
          \ell}\\P^{\Psi<_{\infty}\Psi'}_{\ell}=\limsup_{p\rightarrow\infty}
          P^{\Psi<_{\infty}\Psi'}_{p,\ell} =\lim_{p\rightarrow\infty}
          P^{\Psi<_{\infty}\Psi'}_{p,\ell}\\P_{\Psi<_{\infty}\Psi'}=
          \limsup_{\ell\rightarrow\infty}P^{\Psi<_{\infty}\Psi'}_{\ell}=\lim_{\ell
          \rightarrow\infty}P^{\Psi<_{\infty}\Psi'}_{\ell}.\end{array}
          \end{equation}

           These definitions have some satisfying properties. One is
          that if $\Psi$ and $\Psi'$ are Cauchy sequences then
          exactly one of the following relations is true with
          probability $1$ and the other two are false (true
          with probability $0$):\begin{equation}\label{Psieqltgt}
          \begin{array}{c}\Psi =_{\infty,k,g}\Psi'\\
          \Psi <_{\infty,k,g}\Psi'\\\Psi >_{\infty,k,g}
          \Psi'\end{array}\end{equation}

           This result follows from the observation that $0,1$ are
           the only possible values for $P_{\Psi Re_{i,\infty,k,g}\Psi'}$ for
           $\mathcal{F}_{k}$ valued sequences, $\Psi,\Psi',$ provided
          that $\Psi$ and $\Psi'$ are both Cauchy. That is\begin{equation}
          \label{FRe01}\Psi \mbox{ and } \Psi' \mbox{ are Cauchy}
          \Rightarrow P_{\Psi Re_{i,\infty,k,g}\Psi'} = 0\mbox{ or }1.
          \end{equation}

          To see this it is sufficient to examine $=_{\infty,k,g}$
          as the proofs for the other two relations are similar.  One
          can rewrite Eq. \ref{Ppsinml} in the equivalent form \begin{equation}
          \label{EqPnml} P^{\Psi =_{\infty}\Psi'}_{n,m,\ell}= \sum_{\g,h,s}
          \sum'_{\g',h',s'} |d^{n}_{\g,h,s}|^{2}| f^{m}_{\g',h',s'}|^{2}.
          \end{equation} The prime on the $\g',h',s'$ sum mean that
          the sum is restricted to states $|\g',h',s'\rangle$ that
          are at least as large as $|(g,h,s)-_{A}(+,-\ell)\rangle$
          and no larger than $|(g,h,s)+_{A}
          (+,-\ell)\rangle.$ Since the states $\Psi$ and $\Psi'$
          are Cauchy, the distributions $|d^{n}_{\g,h,s}|^{2}$ and
          $| f^{m}_{\g',h',s'}|^{2}$ become increasingly narrow
          as $n,m$ increase.

          The distributions either lie on top of one another for
          each $\ell$ or they do not. In the first case, for
          sufficiently large $m,n$ the restrictions on the $\g',h',s'$
          sum can be ignored and $\lim_{m,n\rightarrow\infty}
          P^{\Psi =_{\infty}\Psi'}_{n,m,\ell} =1$ for all $\ell$.
          In the second case there is some $\ell$ for which the
          state $|+,-\ell\rangle$ approximately separates the
          distributions. For this and all larger $\ell$ values, the
          overlap probability $P^{\Psi =_{\infty}\Psi'}_{n,m,\ell}$
          in Eq. \ref{EqPnml} approaches $0$ as $m,n\rightarrow\infty.$

          Another useful property of the asymptotic relation
          $=_{\infty}$ is that for each $\mathcal{F}_{k}$
          valued Cauchy sequence $\Psi$ there is a
          $\mathcal{B}_{k,g}$ valued sequence $\Psi'$ such that
          $P^{\Psi=_{\infty}\Psi'}=1$ and $\Psi'$ is Cauchy. The
          definition of $\Psi'$ and proof that $\Psi'$ is Cauchy
          and are summarized here.  The proof that $P^{\Psi=_{\infty}
          \Psi'}=1,$ or that $\Psi=_{\infty}\Psi',$ will not be given
          as it is similar to that for the Cauchy property of $\Psi'.$

          Define $\Psi'(n)=|\g_{n},h_{n},s_{n}\rangle_{k,g}$ to be the state
          that maximizes the probability $P^{\Psi}_{n,l}(\g',h',s')$ where
         \begin{equation}\label{Pmax} \begin{array}{l}P^{\Psi}_{n,l}
         (\g',h',s')=\sum_{\g,h,s}|d^{n}_{\g,h,s}|^{2}: \\ \hspace{1cm}
         ||(\g,h,s)-_{A,k,g}(\g',h',s')|_{A,k,g}\rangle_{k}\leq_{A,k,g}|+,\ell
         \rangle_{k,g}.\end{array}\end{equation} Define $Q^{\Psi}_{n,\ell}$ to be
         that maximum:\begin{equation}\label{defQnl} Q^{\Psi}_{n,\ell}=
         P^{\Psi}_{n,l}(\g_{n},h_{n},s_{n}).\end{equation} Since
         $Q^{\Psi}_{n,\ell}\geq P^{\Psi}_{n,l}(\g',h',s'),$ multiplying
         both sides by $|d^{m}_{\g',h',s'}|^{2}$ and carrying out the
         sum $\sum_{\g',h',s'}$ gives\begin{equation}\label{QgrtrP}
         Q^{\Psi}_{n,\ell}\geq P^{\Psi}_{n,m,l}.\end{equation} Since
         $\Psi$ is Cauchy,\begin{equation}\label{Qlim}\lim_{n\rightarrow
         \infty}Q^{\Psi}_{n,\ell}=1.\end{equation}

         To show that $\Psi'$ is Cauchy, it is sufficient to prove
         that $\lim_{m,n\rightarrow\infty}W_{m,n,\ell}=0$ where
         $$\begin{array}{l}W_{m,n,\ell}=\sum_{\g,h,s}\sum_{\g',h',s'}
         |d^{n}_{\g,h,s}|^{2}|d^{m}_{\g',h',s'}|^{2}:\\ \hspace{1cm}
         ||(\g_{n},h_{n},s_{n})-_{A,k,g}(\g_{m},h_{m},s_{m})|_{A,k,g}\rangle
         >_{A,k,g}|+,-3\ell\rangle_{k,g}.\end{array}$$ If the condition
         is true, then $\lim_{m,n\rightarrow\infty}
         W_{m,n,\ell}=1;$ if it is false, then  $\lim_{m,n\rightarrow\infty}
         W_{m,n,\ell}=0.$ Also $|+,-3\ell\rangle_{k,g}=|+,-\ell
         \rangle_{k,g}+_{A}|+,-\ell\rangle_{k,g}+_{A}|+,-\ell\rangle_{k,g}.$

         Define $X_{m,n,\ell}$ by $$\begin{array}{l}X_{m,n,\ell}=
         \sum_{\g,h,s}\sum_{\g',h',s'}|d^{n}_{\g,h,s}|^{2}
         |d^{m}_{\g',h',s'}|^{2}:\\ \hspace{1cm}
         ||(\g_{n},h_{n},s_{n})-_{A}(\g,h,s)|_{A}\rangle+_{A}||(\g,h,s)-_{A}
         (\g',h',s')|_{A}\rangle \\ \hspace{2cm}+_{A}||(\g',h',s')-_{A}
         (\g_{m},h_{m},s_{m})|_{A}\rangle >_{A}|+,-3\ell\rangle_{k,g}.
         \end{array}$$ (Subscripts $k,g$ are suppressed here.) Since
         $$\begin{array}{l}||(\g_{n},h_{n},s_{n})-_{A}(\g_{m},h_{m},g_{m})
         |_{A}\rangle \\ \hspace{1cm}\leq_{A}||(\g_{n},h_{n},s_{n})-_{A}
         (\g,h,s)|_{A}\rangle+_{A}||(\g,h,s)-_{A}(\g',h',s')|_{A}\rangle
         \\ \hspace{2cm}+_{A}||(\g',h',s')-_{A}(\g_{m},h_{m},s_{m})|_{A}
         \rangle\end{array},$$ one has the result that \begin{equation}
         \label{WX}W_{m,n,\ell}\leq X_{m,n,\ell}.\end{equation}

         The condition in the definition of $X_{m,n,\ell}$ is
         satisfied only if at least one of the component states is
         $\geq_{A}|+,-\ell\rangle.$ If this holds for the first or
         third component, then $X_{m,n,\ell}\leq 1-Q_{n,\ell}$ or
         $X_{m,n,\ell}\leq 1-Q_{m,\ell}.$ If it holds for the
         second component, then $X_{m,n,\ell}\leq 1-P^{\Psi}_{m,\ell}.$

          Eq. \ref{Qlim} and the fact that $\Psi$ is Cauchy gives the result
          that $\lim_{m,n\rightarrow\infty}X_{m,n,\ell}=0.$ It
          follows from Eq. \ref{WX} that $\lim_{m,n\rightarrow\infty}
          W_{m,n,\ell}=0.$  This gives the final result that
          $|(\g_{m},h_{m},s_{m})-_{A}(\g_{n},h_{n},s_{n})|_{A}
          \rangle\leq_{A}|+,-\ell\rangle$ and thus that $\Psi'$ is
          Cauchy.

         \subsection{Basic Operations on State Sequences}
         \label{BOSS}
         The problems requiring the definition of asymptotic
         relations do not appear to be present in the definition of
         basic relations on the sequences. For $\mathcal{B}_{k,g}$
         valued sequences $\Psi$ and $\Psi'$ one uses Eqs.
         \ref{arithops} and \ref{addnplA} to define $O_{\nu,S,k,g}$
         by\begin{equation}\label{defOseq}O_{\nu,S,k,g}\Psi\Psi'=\Theta
         \end{equation} where for $\nu =1,2,3$ and each $n$,\begin{equation}
         \label{OSOA}\Theta(n)=\Psi(n)\times\Psi'(n)\times\Psi''(n)\end{equation}
         and \begin{equation}\label{psiOp}\Psi''(n)=\Psi(n)O_{\nu,A,k,g}
         \Psi'(n).\end{equation}  The product structure of the
         elements of $\Theta$ allows one to write\begin{equation}
         \label{thetaprod}\Theta =\Psi\Psi'\Psi''\end{equation}
         as the product of $3$ state sequences.

         For $\nu =4$  one division operator,
         $\div_{S,k,g},$ can be defined as an operator that is diagonal in the
         infinite number of state division operators, $\div_{A,\ell}.$
         One has\begin{equation}\label{divS}\div_{S,k,g}\Psi\Psi'=
         \Psi\Psi'\Psi''\end{equation}where\begin{equation}\label{divSdivA}
         \Psi''(n)=\Psi(n)\div_{A,k,g,n}\Psi'(n).\end{equation} Note the
         subscript $n$ in $\div_{A,k,g,n}.$

         Application of this definition to more general
         $\mathcal{F}_{k}$ valued sequences $\Psi$ and $\Psi'$
         generates a single sequence $\Theta$ of entangled states
         that cannot be represented in the product form of Eq.
         \ref{thetaprod}.  From Eqs. \ref{Opsipsip} and \ref{dfpsi},
         one has\begin{equation}\label{Opsilinsup}O_{\nu,A,k,g}
         \Psi\Psi'=\Theta\end{equation} where\begin{equation}
         \label{Thetaentngl}\begin{array}{l}\Theta(n)=\sum_{\g,h,s}
         \sum_{\g',h',s'}|d^{n}_{\g,h,s}|^{2}|f^{n}_{\g',h',s'}|^{2}
         \\\hspace{1cm}|\g,h,s\rangle|\g',h',s'\rangle|(\g,h,s)
         O_{\nu,A,k,g}(\g',h',s')\rangle.\end{array}\end{equation}

         As shown, $\Theta$ is not a result sequence in the sense that
         $\Psi''$ was. To obtain a result sequence one must take the
         trace over the two initial states for each element of $\Theta.$
         This gives a sequence $\mathcal{P}_{\Psi,\Psi}$ of
         density operators where\begin{equation}\label{rhotheta}
         \begin{array}{l}\mathcal{P}_{\Psi,\Psi}(n)=\sum_{\g,h,s}
         \sum_{\g',h',s'}|d^{n}_{\g,h,s}|^{2}|f^{n}_{\g',h',s'}|^{2}
         \\ \hspace{1cm}\times\rho_{|(\g,h,s)O_{\nu,A,k,g}(\g',h',s')
         \rangle}.\end{array}\end{equation}

         Inclusion of these sequences into the definitions presented
         so far requires expansion of the material to
         define the Cauchy condition and asymptotic relations for
         sequences of density operators. Since this has not yet
         been done, this will be left to future work.  Also it is
         not clear if element definitions of $O_{\nu,S,k,g},$ as
         is done in Eq. \ref{OSOA}, are useful here. In any case,
         one can proceed without these extensions.  Also the main
         results are not affected by this lack.

         \subsection{Quantum Representation of Real Numbers} \label{QRRN1}

         The asymptotic equality relation $=_{\infty,k,g}$ can be used
         to define equivalence classes of Cauchy sequences. Two
         sequences $\Psi$ and $\Psi'$ are equivalent if and only if
         they are asymptotically equal:\begin{equation}\label{equiveq}
         \Psi\equiv\Psi'\Leftrightarrow \Psi=_{\infty,k,g}\Psi'
         \end{equation} It is straightforward to show from the
         properties of $=_{\infty,k,g}$ that $\equiv$ has the right
         properties for a definition of equivalence.

         For each Cauchy sequence $\Psi$ let $[\Psi]$ denote the equivalence
         class containing $\Psi.$ As might be guessed, the set of all these
         equivalence classes is a quantum representation of the real numbers.
         Let $R_{k,g}$ denote the set. The subscripts $k,g$ indicate that the
         representation depends on both the base $k$ and the gauge or basis choice
         $g$.

         As has been seen  each class $[\Psi]$ contains many $\mathcal{F}_{k}$
         valued sequences and at least one $\mathcal{B}_{k,g}$ valued
         sequence. From this one concludes that the quantum
         equivalence classes are larger than the classical
         equivalence classes but that no new classes are present.

         The basic relations and operations can be lifted from
         sequences to the equivalence classes to define the basic
         relations and operations for a real number field. Let
         $Re_{i,R,k,g}$ denote the two relations $=_{R,k,g}$ for
         $i=1$ and $<_{R,k,g}$ for $i=2.$ Let $[\Psi]$ and $[\Psi']$
         denote two equivalence classes of Cauchy sequences.  Then
         \begin{equation}\label{defEqR}[\Psi]Re_{i,R,k,g}[\Psi']
         \Leftrightarrow\Psi Re_{i,\infty,k,g}\Psi'. \end{equation}
         This definition holds for all $\mathcal{F}_{k}$ valued
         sequences.

         The field operations, $O_{\nu,R,k,g}$ for $\nu=1-4$
         $(+,\times,-,\div),$ can be defined on equivalence
         classes through their definitions on $\mathcal{B}_{k,g}$
         valued sequences. Let $\Psi,\Psi',\Psi''$ be Cauchy
         sequences that satisfy Eqs. \ref{defOseq} and \ref{thetaprod}.
         Define $O_{\nu,R,k,g}$ by\begin{equation}\label{DefOR}
         [\Psi] O_{\nu,R,k,g}[\Psi'] =[\Psi''].\end{equation} This
         use of $\mathcal{B}_{k,g}$ valued sequences to define
         the field operations definition is done only because
         Cauchy sequences of density operators are not included
         here. Their inclusion would allow direct definitions of
         the field operators for all Cauchy sequences.

         To verify that $R_{k,g}$ is a representation of the
         real numbers, one must show that $R_{k,g}$ and the relations,
         $Re_{i,R,k,g},$ and operations $O_{\nu,R,k,g},$ satisfy the
         real number axioms of a complete ordered field
         \cite{Hewitt}. Some details of this were given in
         \cite{BenRRCNQT}, so it will not be repeated here.
         The proof is, in many ways, similar to that given for
         the usual classical Cauchy sequences of rational
         numbers \cite{Burkhill}.

          \section{Space of Real Number Representations and
          Associated Transformations} \label{SRNRAT} As described,
          the quantum theory representations of real numbers,
          $R_{k,g},$ depend on a base $k$ and a gauge
           $g.$ Recall that $k$ denotes the dimensionality of
           the Hilbert space of states for each single $q_{k}$ and
           $g$ denotes a gauge field of basis sets on $I\times I,$ Eq.
           \ref{gjhB}.

           For each pair $k,g$ one has a quantum representation
           $R_{k,g}$ of the real numbers. Any pair, $R_{k,g},R_{k',g'}$
           of real number representations are isomorphic as all
           representations of the real numbers (axiomatized by
           second order axioms) are isomorphic \cite{Barwise}.
           However this does not mean that they are identical.  For
          instance, Cauchy sequences of $q_{k}$ string states, which
          are elements of the equivalence classes in $R_{k,g},$ are
          distinct from Cauchy sequences of $q_{k'}$ string states,
          which are elements of equivalence classes in $R_{k',g}$,
          as $q_{k}$ and $q_{k'}$ systems are different.

          Similarly $\mathcal{B}_{k,g}$ valued sequences  $\Psi$
          of $q_{k}$ string states are different from
          $\mathcal{B}_{k,g'}$ valued sequences.  Also the
          definition of the Cauchy condition, Eq. \ref{cauchy} is
          both $k$ and $g$ dependent. These dependencies can
          be seen from Eqs. \ref{UBg}-\ref{Uexpand} which show the
          relations between the two basis sets and between the single
          $q_{k}$ A-C operators for each of the two basis sets.

          These considerations show that the set of all
          representations $R_{k,g}$ can be regarded as a space of
          representations parameterized by a two dimensional space
          of all pairs, $k,g$. Transformations $k,g\rightarrow
          k',g'$ induce transformations $R_{k,g}\rightarrow
          R_{k',g'}$ on the representation space.  The components of
          the transformations on the representation space are operators
          that change bases, $R_{k,g}\rightarrow R_{k',g}$, and
          operators that change the gauge, $R_{k,g}\rightarrow R_{k,g'}.$

          \subsection{Gauge Changing Operators}\label{GCO}Gauge
          changing operators that act on sequences can be defined
          from the gauge transformations, $U_{k}$ as defined in Eqs.
          \ref{UBg} and Eq. \ref{Urast}. To achieve this, let $\Psi$
          and $\Psi'$ be respective $\mathcal{B}_{k,g}$ and
          $\mathcal{B}_{k,g'}$ valued sequences where\begin{equation}
          \label{PsiPsipG}\begin{array}{c}\Psi(n)=|\g_{n},h_{n},s_{n}
          \rangle_{k,g}\\ \Psi'(n)=|\g_{n},h_{n},s_{n}\rangle_{k,g'}.
          \end{array}\end{equation} Define the
          operator, $\mathcal{U}_{k},$ by \begin{equation}
          \label{Umathcal}\Psi'= \mathcal{U}_{k} \Psi\end{equation} where
          \begin{equation}\label{UkPsin}|\g_{n},h_{n},s_{n}
          \rangle_{k,g'}=U_{k}|\g_{n},h_{n},s_{n}\rangle_{k,g}.
          \end{equation} Here $g$ and $g'$ are related by
          \begin{equation}\label{gpUkg} g'=U_{k}g.\end{equation}

          This shows that the elements of $\Psi'$, $\Psi'(n),$ are
          the same states, relative to the transformed basis as the
          elements ,$\Psi(n)$, of $\Psi$ are, relative to
          the original basis.  However, relative to the original basis,
          the states $\Psi'(n)$ are different from the states $\Psi(n).$
          This can be seen by expanding the states $\Psi'(n)$ in terms of
          the original basis.

          The definition of $\mathcal{U}_{k}$ extends by linearity
          to $\mathcal{F}_{k}$ valued sequences. If $\Psi$ is such a
          sequence where\begin{equation}\label{Psinexpnd}
          \Psi(n)=\sum_{\g,h,s}d^{n}_{\g,h,s}|\g,h,s\rangle_{k,g},
          \end{equation} then $\Psi'$ is related to $\Psi$ by Eq.
          \ref{Umathcal} where \begin{equation}\label{Psinexpnd1}
          \Psi'(n)=\sum_{\g,h,s}d^{n}_{\g,h,s}|\g,h,s\rangle_{k,g'}
          \end{equation} Note the replacement of $g$ by $g'$ on the
          right hand side.

          The definition of $\mathcal{U}_{k}$ can be lifted to apply
          to equivalence classes of Cauchy sequences to relate
          $R_{k,g}$ to $R_{k,g'}$ as in $R_{k,g'}=\mathcal{U}_{k}
          R_{k,g}.$ The validity of this depends on the preservation
          of the Cauchy property under the action of $U_{k}.$  That
          is, if $\Psi$ is a g-Cauchy sequence, then $\Psi'=\mathcal{U}_{k}\Psi$
          is a $g'-$Cauchy sequence.\footnote{The $g'-$Cauchy condition is given
          by Eq. \ref{cauchy} with the subscript $g'$ replacing $g$.} To show that
          this is the case one has to define the g'-Cauchy condition relative to the
          basis states in $\mathcal{B}_{k,g}.$  This is\begin{equation}\label{Ucauchy}
          \begin{array}{c}\forall{\ell}\exists{p}\forall{j,m}>p |(|U_{k}
          (\gamma_{j},h_{j},s_{j})-_{A,k,g'} U_{k}(\gamma_{m},h_{m},s_{m})
          |_{A,k,g'})\rangle_{k,g'}\\\leq_{A,k,g'}U_{k}|+,-\ell\rangle_{k,g}.
          \end{array}\end{equation} Here $U_{k}|\g_{j,},h_{j},s_{j}\rangle_{k,g}
          =|\g_{j},h_{j},s_{j}\rangle_{k,g'}$, $\leq_{A,k,g'}=U_{k}\leq_{A,k,g}
          U^{\dag}_{k},$ and $-_{A,k,g'}=U_{k}\times U_{k}\times
          U_{k}-_{A,k,g}U^{\dag}_{k}\times U^{\dag}_{k}.$ It is a
          straightforward exercise to show that, for this definition,
          the Cauchy property is preserved under the action of $U_{k}.$

          The definition of $\mathcal{U}_{k}$ shows that these
          operators form a group of transformations. If $\mathcal{U}_{k}$
          and $\mathcal{U}'_{k}$ are gauge  transformations, for
          Cauchy sequences, or for equivalence classes in $R_{k,g}$, then
          so is their product $\mathcal{U}_{k}\mathcal{U}'_{k}$. Also each
          $\mathcal{U}_{k}$  has an inverse $\mathcal{U}^{-1}_{k}$.  The
          group property follows from the fact that the $U_{k},$ on which
          the $\mathcal{U}_{k}$ are based, are products of elements of the
          unitary group $U(k).$

          \subsection{Base Changing Operators}\label{BCO}

          One would like to describe the base changing transformations
          for Cauchy sequences by lifting the base changing transformations
          $\tilde{W}_{k',k}$ for the $q_{k}$ string states to transformations
          on the Cauchy sequences. One  first thinks of doing this by defining an
          operator $\mathcal{W}_{k',k}$ on $\mathcal{B}_{k,g}$ valued sequences
          $\Psi$. One would set \begin{equation}\label{Psikkp} \Psi'=
          \mathcal{W}_{k',k}\Psi.\end{equation} Here $\Psi'$ is a
          $\mathcal{B}_{k',g}$ valued sequence such that for each $n$
          \begin{equation}\label{basech}\Psi'(n)=|\g_{n},h_{n},s'_{n}\rangle_{k',g}=
         \tilde{W}_{k',k}|\g_{n},h_{n},s_{n}\rangle_{k,g}=\tilde{W}_{k',k}\Psi(n)
         \end{equation}

         The problem with this definition is that the domain and range
         of $\tilde{W}_{k',k}$ depend on the relation of the prime
         factors of $k$ and $k'.$ If $k$ and $k'$ are relatively
         prime, then this definition fails as $\tilde{W}_{k',k}$ is
         not defined on any of the non integer states.

         One way around this impasse is to generalize the definition
         of $\tilde{W}_{k',k}$ to operators $\tilde{W}_{k',k,\ell}$
         for different nonnegative integers $\ell.$ Here\begin{equation}
         \label{kpWkell}|\g_{n},h_{n},s'_{n}\rangle_{k',g}=
         \tilde{W}_{k',k,\ell}|\g_{n},h_{n},s_{n}\rangle_{k,g}
         \end{equation} is a base $k'$ state that represents the
         same number as the base $k$ state $|\g_{n},h_{n},s_{n}
         \rangle_{k,g}$ \emph{up to accuracy $|+,-\ell\rangle_{k',g}$,}.
         This removes the problem because, for each $\ell$,
         $\tilde{W}_{k',k,\ell}$ is defined on all $q_{k}$ string states
         in $\mathcal{F}_{k}.$ Also $\tilde{W}_{k',k,\ell}=
         \tilde{W}_{k',k}$ on the integer state subspace of
         $\mathcal{F}_{k}.$

         The desired definition of $\mathcal{W}_{k',k}$ is that it be an
         isomorphism from $R_{k,g}$ to $R_{k',g}.$  This is
         equivalent to requiring that $\mathcal{W}_{k',k}\Psi$ belongs to
         the equivalence class in $R_{k',g}$ that represents
         the same number  as the equivalence class in $R_{k,g}$
         that contains $\Psi.$ A proposed method of achieving this is
         by  a definition that is diagonal in $n$ and in $\ell.$

         To this end one defines $\mathcal{W}_{k',k}$  by replacing
         $\tilde{W}_{k',k}$ with $\tilde{W}_{k',k,n}$ in Eq. \ref{basech}
         to get \begin{equation}\label{basechell} \Psi'(n)=|\g_{n},
         h_{n},s'_{n}\rangle_{k',g}=\tilde{W}_{k',k,n}|\g_{n},h_{n},s_{n}
         \rangle_{k,g}=\tilde{W}_{k',k,n}\Psi(n).\end{equation}

         The operator $\mathcal{W}_{k',k}$ must satisfy two properties:
         The sequence $\Psi=\mathcal{W}_{k',k}\Psi$ must be Cauchy if
         $\Psi$ is Cauchy, and the two sequences, $\Psi'$ in $R_{k',g}$
         and $\Psi$ in $R_{k,g}$, must represent the same real number.
         (Here and in the following, Cauchy sequences will often be
         stand ins for equivalence classes of the sequences.) An equivalent
         requirement is that $\mathcal{W}_{k',k}$ is an isomorphism from
         $R_{k,g}$ to $R_{k',g}.$ It preserves the basic field relations of
         equality and ordering and the operations of addition, multiplication
         and their inverses.

          \section{Quantum Representations of Complex
          Numbers}\label{QRCN}
          The simplest path to  descriptions of quantum representations of
          complex numbers is their representation as ordered pairs
          of real number representations.  If $[\Psi]$ and $[\Psi']$
          represent two quantum real numbers, then $([\Psi],[\Psi'])$
          represents a quantum complex number where
          $[\Psi]=[\Psi]^{r}$ and $[\Psi']=[\Psi']^{i}$ represent
          the real and imaginary components. The basic field
          relations $=_{R,k,g},<_{R,k,g}$ and operations
          $+_{R,k,g},\times_{R,k,g},-_{R,k,g},\div_{R,k,g}$ would be
          extended to relations $=_{C,k,g},<_{C,k,g}$ and operations
          $+_{C,k,g},\times_{C,k,g},-_{C,k,g},\div_{C,k,g}$
          following the standard rules.

          The rest of this section can be skipped over by readers
          using the above definitions. The following development
          is based on the observation that all physical representations
          of complex numbers, such as in computations, are by ordered
          pairs of single string representations of rational numbers.
          This corresponds here to extending the treatment of rational number
          representations, as states of $q_{k}$ strings, to ordered
          pairs of states of $q_{k}$ strings. These represent the real
          and imaginary components of complex rational numbers.
          Application of the Cauchy condition separately to
          the real and imaginary components gives a description
          of Cauchy sequences of complex rational number
          representations.  This gives quantum representations of
          complex numbers as equivalence classes of these Cauchy sequences.

          One way to proceed is to continue working with one type of
          qukit but increase the number of sign qubit types from one to
          two.\footnote{This differs from the approach in \cite{BenRRCNQT}
          which uses two types of  qukits and qubits.}. The  two qubit
          types are represented by A-C operators $\cd_{\g,0,h},d^{\dag}_{\delta,0,h}$
          and their complex conjugates.  Here $\cd_{\g,0,h}$ and
          $d^{\dag}_{\delta,0,h}$ represent sign creation operators for the
          real and imaginary number components where $\g=+,-$ and
          $\delta =+i,-i$ at site $0,h'$

          Complex rational numbers are represented here by pairs of qukit
          string states, $|h,\g,s;h',\delta,t\rangle_{k,g},$ with different
          $h$ values. In terms of A-C operators one has\begin{equation}
          \label{complxst}|h,\g,s;h_{1},\delta,t\rangle_{k,g}=\cd_{\g,0,h}
          \ads_{[(l,h),(u,h)]};\dd_{\delta,0,h_{1}}\adt_{[(l',h_{1}),
          (u',h_{1})]}|0\rangle.\end{equation} The state $|0\rangle$ denotes the qukit
          vacuum and $0,h$ and $0,h_{1}$ denote the locations of the sign
          qubits. As before $[(l,h),(u,h)]$ and $[(l',h_{1}),(u',h_{1})]$ denote
          lattice intervals where $l\leq 0\leq u$ and $l'\leq 0\leq u'$.   Also
         \begin{equation}\label{asat}\begin{array}{c}\ads_{[(l,h),(u,h)]}
          =\ad_{s(u),u,h}\ad_{s(u-1),u-1,h}\cdots\ad_{s(l),l,h}\\
          \adt_{[(l',h_{1}),(u',h_{1})]} =\ad_{t(u'),u',h_{1}}\ad_{t(u'-1),
          u'-1,h_{1}}\cdots\ad_{t(l'),(l',h_{1})}\end{array}
         \end{equation} where $s$ and $t$ are $0,\cdots,k-1$ valued functions
         with integer interval domains $[l,u]$ and $[l',u']$
         respectively. The subscript $g$ denotes the implicit gauge
         choice for the  $q^{r}_{k}$ string states at each site of
         $I\times I.$

         A consequence of this representation is that if one has
         many pairs of string states, they are expressed in the
         A-C formalism as one long string of creation operators
         acting on $|0\rangle.$ One then needs a method of
         determining the association between the imaginary and
         real strings.  One of the different ways to do this is to
         describe the pairs as those in which $h$ is close or next
         to $h_{1}$. Here some method will be assumed implicitly as
         which one is used does not affect the results obtained in
         this paper.

         The definitions of arithmetic relations and operations
         given for states of $q_{k}$ strings can be easily
         extended to states of pairs  of $q_{k}$ strings
         following the usual arithmetic rules for operations on
         complex numbers. For arithmetic equality one has
         \begin{equation}\label{ReAck}
         \begin{array}{l}|h,\g,s;h_{1},\g_{1},t\rangle_{k,g}=_{c,k,g}
         |h',\g',s',h'_{1},\g_{1}',t'\rangle_{k,g} \\ \Leftrightarrow
         ( |h,\g,s\rangle_{k,g}=_{r,k,g}|h',\g',s'\rangle_{k,g}
         \\ \mbox{and }|h_{1},\g_{1},t\rangle_{k,g}=_{i,k,g}|h'_{1},\g'_{1},t'
         \rangle_{k,g}).\end{array}\end{equation} Ordering relations
         are usually not considered because they are only partly
         defined (complex numbers cannot be ordered). The $c,r,i$ in
         the subscripts denote complex, real, and imaginary,
         respectively.

         For the operations let $\tilde{O}_{c,k,g}$ be a unitary
         operator denoting any of the four operations $+_{c,k,g},
         \times_{c,k,g},-_{c,k,g},\div_{c,k,g,\ell}.$ The action of
         any of these on complex rational states can be represented by
         \begin{equation}\label{defO}\begin{array}{l}\tilde{O}_{c,k,g}
         |h,\g,s;h_{1},\delta,t\rangle_{k,g} |h',\g',s';h'_{1},\delta',t\rangle_{k,g} \\
         =|h,\g,s;h_{1},\delta,t\rangle_{k,g} |h',\g',s';h'_{1},\delta',t'\rangle_{k,g}
         |h'',\g'',,s'';h''_{1},\delta'',t''\rangle_{k,g}\end{array}\end{equation}
         where\begin{equation}\label{stOAstp}|\asatpp\rangle_{k,g}
         =_{c,k,g}|(\asat)O_{c,k,g}(\asatp)\rangle_{k,g}.\end{equation}
        The expression $|(\asat)O_{c,k,g}(\asatp)\rangle_{k,g}$ with
         $O$ inside $|-,-\rangle$  represents the rational
        string state resulting from carrying out the operation
        $O_{c,k,g}$. Unitarity is satisfied by preserving the two input
        states and creating a result state.

         The arithmetic operations create entangled states when
         applied to linear superpositions of the basis states.
         One has \begin{equation}\label{Oentngl}\begin{array}{l}
         \tilde{O}_{c,k,g}\psi\psi^{\p} = \sum_{h,\g,s,h_{1},\delta,t}
         \sum_{h',\g', s',h'_{1},\delta',\tp} {k,g}_\langle\asat|\psi\rangle
         \\\hspace{0.5cm}\times {k,g}_\langle\asatp|\psi^{\p}\rangle
         |\asat\rangle_{k,g}|\asatp\rangle_{k,g}\\
        \hspace{1cm}\times|(\asat)O_{c,k,g}(\asatp)\rangle_{k,g}.
        \end{array}\end{equation} Taking the trace over the $\psi$
        and $\psi^{\p}$ component states gives a mixed state
        \begin{equation}\label{rhoplus}\begin{array}{l}\rho_{\psi
        O_{c,k,g}\psi'}=\sum_{h,\g,s,h_{1},\delta,t}\sum_{h',\g',
        s',h'_{1},\delta',\tp}|\langle\asat|\psi\rangle|^{2}\\
        \hspace{.5cm}\times|\langle\asatp|\psi^{\p}\rangle|^{2}
        \rho_{(\asat)O_{c,k,g}(\asatp)}\end{array}\end{equation}
        that represents the result of the operation.

        Determination of the exact form of the state
        $|\asatpp\rangle_{k,g}$ from Eq. \ref{stOAstp} for the
        different arithmetic operations is somewhat lengthy,
        but straightforward.  It involves translation of the
        usual rules for implementation of arithmetic operations
        on complex numbers into those on quantum states. For
        example, for multiplication one uses the relations
        \begin{equation}\label{cmplxtimes}\begin{array}{c}
        \dd_{\g,0,h}\times\dd_{\g',0,h'}=\cd_{\g'',0,h''}
        \mbox{ where }\g''=+,[-]\mbox{ if }\g\neq [=]\g'\\
        \cd_{\g,0,h}\times\dd_{\g',0,h'}=\dd_{\g'',0,h''}
        \mbox{ where }\g''=\g',[\g'\neq\g'']\mbox{ if }\g=+
        [\g=-]\\\cd_{\g,0,h}\times\cd_{\g',0,h'}=\cd_{\g'',0,
        h''} \mbox{ where }\g''=+,[-]\mbox{ if }\g=\g'
        [\g\neq\g'].\end{array}\end{equation}

        Quantum representations of complex numbers are based on
        application of the Cauchy condition  to the real and
        imaginary components separately of a sequence of states of $q_{k}$
        string pairs. The sequence $\Psi$ where
        $\Psi(n)=|h_{n},\g_{n},s_{n};h'_{n},\delta_{n},t_{n}\rangle_{k,g}$ of
        states is a Cauchy sequence if the following is satisfied:
        \begin{equation}\label{cauchyri}\begin{array}{l}\forall{\ell}
        \exists{p}\forall{j,m>p}||(h_{j},\g_{j},s_{j})
        -_{r,k,g}(h_{m},\g_{m},s_{m})|_{r,k,g}\rangle_{k,g}
        <_{r,k,g}|+,-\ell\rangle_{k,g}\\\mbox{and }||(h'_{j},\delta_{j},t_{j})
        -_{i,k,g}(h'_{m},\delta_{m},t_{m})|_{i,k,g}\rangle_{k,g}
        <_{r,k,g}|+,-\ell\rangle_{k,g}.\end{array}
        \end{equation} Here $|+,-\ell\rangle_{k,g}$ is the
        state corresponding to the number $k^{-\ell}.$

          Extension of the Cauchy condition to sequences of linear
          superpositions of complex rational string states is
          similar to that for sequences of superpositions of real
          rational states.  Such a sequence is Cauchy if the
          probability is unity that both the real and imaginary
          components satisfy the Cauchy condition.

          The definition of equivalence for the real number representations
          extends here to complex number representations.  Two Cauchy
          sequences $\Psi$ and $\Psi'$ are equivalent if the real and
          imaginary components of $\Psi$ and $\Psi'$ are asymptotically
          equal. Let $\Psi$ and $\Psi'$ be $\mathcal{B}_{k,g}$ valued
          sequences where for each $n$\begin{equation}\label{PsiPsipC}
          \begin{array}{c}\Psi(n)=|h_{n},\g_{n}, s_{n};h_{1,n},\delta_{n},t_{n}
          \rangle_{k,g}\\ \Psi'(n)= |h'_{n},\g'_{n},s'_{n};h'_{1,n},\delta'_{n},
          t'_{n}\rangle_{k,g}.\end{array}\end{equation} Then
          \begin{equation}\label{EqasympC}\begin{array}{l}\Psi
          =_{\infty,S,k,g}\Psi'\mbox{ if }\forall{\ell}\exists{p}
          \forall{j,m>p} \\\hspace{0.5cm}||(h_{j},\g_{j},s_{j})
          -_{r,k,g}(h'_{m},\g'_{m},s'_{m})|_{r,k,g}
          \rangle_{k,g}\leq_{r,k,g}|+,-\ell|\rangle_{k,g}\mbox{ and}
          \\\hspace{0.5cm}||(h_{1,j},\delta_{j},t_{j})
          -_{i,k,g}(h'_{1,m},\delta'_{m},t'_{m})|_{i,k,g}
          \rangle_{k,g}\leq_{r,k,g}|+,-\ell|\rangle_{k,g}.
          \end{array}\end{equation}  From this definition one
          has\footnote{It is
          easy to see that this definition of $\equiv$ has the
          necessary properties of symmetry, reflexivity, and
          transitivity. These follow from the corresponding
          properties of $=_{\infty,S,k,g}.$}\begin{equation}
          \label{equivdef}\Psi\equiv\Psi'\mbox{ if }
          \Psi=_{\infty,S,k,g}\Psi'.\end{equation}

          The set $C_{k,g}$ of complex numbers is defined to be the
          set of equivalence classes $[\Psi]$ where $\Psi$ is a
          Cauchy sequence of $q^{r}_{2}q_{k},q^{i}_{2}q_{k}$
          string pairs. Here $q^{r}_{2}$ and $q^{i}_{2}$ denote the
          real and imaginary sign qubits.
          As was the case for $R_{k,g},$ each equivalence class
          is larger than the corresponding classical equivalence
          class, but there are no new equivalence classes.  This
          follows from the observation that each class contains
          at least one $\mathcal{B}_{k,g}$ valued sequence.

          The basic field relation $=_{C,k,g}$ is defined by
          \begin{equation}\label{EqEquivcl}[\Psi]=_{C,k,g}[\Psi']
          \mbox{ if }\Psi=_{\infty,S,k,g}\Psi'\end{equation}
           The operations $\tilde{+}_{C,k,g},\tilde{-}_{C,k,g},
          \tilde{\times}_{C,k,g},\div_{C,k,g},$ are defined in a
          similar fashion.  For $\mathcal{B}_{k,g}$ valued Cauchy
          sequences one has expressions similar to  Eqs.
          \ref{defOseq} et seq:\begin{equation}\label{Cops}
          \tilde{O}_{\nu,C,k,g}[\Psi][\Psi']=[\Psi][\Psi'][\Psi'']
          \end{equation} Here $\tilde{O}_{\nu,C,k,g}$ with $\nu
          =1,2,3,4$ is a stand in for the four operations. For $\nu
          =1,2,3$ the class $[\Psi'']$ contains all Cauchy sequences
          asymptotically equal to $\Psi''$ where\begin{equation}
          \label{OnuCkg}\Psi''(n)=|\Psi(n)O_{\nu,c,k,g}\Psi'(n)
          \rangle_{k,g}.\end{equation}  For $\nu=4$,
          $(\div_{C,k,g})$ one has a diagonal definition similar
          to Eq. \ref{divSdivA}\footnote{The specific
          definitions of these operations follows those for complex
          numbers.  As examples, for multiplication, if $\Psi(n)=
          |h_{n},\g_{n},s_{n};h_{1,n},\delta_{n},t_{n}\rangle_{k,g}=|x,iy\rangle$ and
          $\Psi'(n)=|h'_{n},\g'_{n},s'_{n};h'_{1,n},\delta'_{n},t'_{n}\rangle_{k,g}
          =|x',iy'\rangle$, then $$\begin{array}{l}\Psi''(n)=
          |(x\times_{r,k,g}x')-_{r,k,g}(y\times_{i,k,g}y');
          \\\hspace{1cm}(x\times_{i,k,g}y')+_{i,k,g}(x'
          \times_{i,k,g}y)\rangle_{k,g}.\end{array}$$ Division
          to accuracy $n$ of $|x,y\rangle$ by $|x',y'\rangle$ is
          done by carrying out the division to accuracy $n$
          indicated by $$|x'',y''\rangle = |[Re,Im]\div_{c,k,g,n}(x'\times
          x')+(y'\times y')\rangle$$ where $Re=x\times x'+y\times y'$
          and $Im =x'\times y-x\times y'.$}:\begin{equation}
          \label{OdivCkg}\Psi''(n)=|\Psi(n)\div_{c,k,g,n}\Psi'(n)
          \rangle_{k,g}.\end{equation}

          As is the case for real number representations, these
          relations and operations extend to $\mathcal{F}_{k}$ valued
          Cauchy sequences. Details will not be given here as they are
          an extension of those for the real number representations.

          \section{Fields of Quantum Reference Frames}\label{FQRF}
          At this point it is good to step back and view some
          consequences of the existence of the many different
          representations of $R$ and $C$. All physical theories
          considered to date, and many mathematical theories, can be
          regarded as theories that are  based on the real and
          complex numbers.  Included are quantum and classical
          mechanics, quantum field theory, QED, QCD, special and
          general relativity, and string theory.  It follows that
          for each representation $R_{k,g},C_{k,g}$ of $R$ and $C$ one has
          a corresponding representation of physical theories as
          mathematical structures based on $R_{k,g},C_{k,g}$

         The large number of theories based on $R,C$ suggests that
         one associate a reference frame $F_{k,g}$ with each $R,C$
         representation, $R_{k,g},C_{k,g}$.  Here $R_{k,g},C_{k,g}$ is
         referred to as the base of frame $F_{k,g}$. The frame $F_{k,g}$
         contains representations of all physical theories that are
         representable as structures based on $R_{k,g},C_{k,g}$.

         The large number of real and complex number representations
         and associated reference frames suggests that one define a
         frame field $F$ over the two dimensional parameter space
         $\{k,g\}.$ The components of $F$, as a map from  $\{k,g\}$
         to a set of reference frames, are the frames $F_{k,g}$ at each
         value $k,g.$  Note that the parameter $g$ is unique to
         quantum theory representations as it is not applicable to
         representations based on states of classical kit strings.
         However, the parameter $k$ is common to both qukit and
         kit string representations.

\begin{figure}[ht]\begin{center}
         \resizebox{200pt}{200pt}{\includegraphics[280pt,160pt]
         [620pt,520pt]{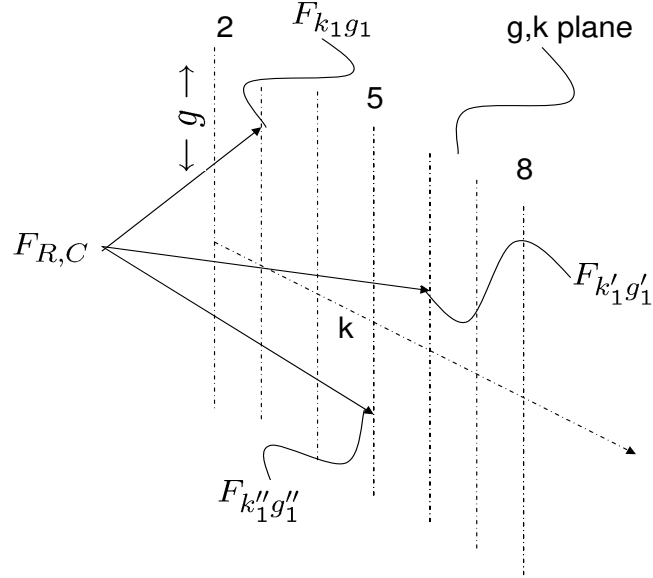}}\end{center}
         \caption{Schematic illustration of frames coming from frame
         $F_{R,C}.$ The frames are based on quantum representations of real and
         complex numbers in $F_{R,C}.$ The distinct vertical lines
         in the $k,g$ plane denote the discreteness of the integral
         values of $k\geq 2.$ Only three of the infinitely many
         frames coming from $F_{R,C}$ are shown. Here $k$ denotes
         the qukit base and $g$ denotes a gauge or basis choice.}
         \label{FIQ1} \end{figure}

         This construction is shown schematically in Figure \ref{FIQ1}
         for three of the infinite number of values of $k,g$. This
         is shown by solid arrows coming from the parent frame
         $F_{R,C}$ to three of the infinitely many descendent frames.

         This use of reference frames has much in common with other
         uses of reference frames in physics and particularly in
         quantum theory \cite{Aharonov,Bartlett,vanEnk,Enk,Poulin,Bartlett1}.
         In special relativity, inertial coordinate systems define
         reference frames for describing physical dynamics. In quantum
         cryptography, Alice and Bob pick a polarization direction
         to define a reference frame for sending messages encoded in
         qubit string states. Here each reference frame carries
         representations of all physical theories as mathematical
         structures based on the real and complex number base of the
         frame.

         It is of interest to examine what observers can and cannot
         see in the different frames. To begin it is assumed that an
         observer $O_{R}$ in the parent frame $F_{R,C}$ regards the
         real and complex numbers in the frame base  as elementary
         objects. The only  relevant properties they have are those
         required by the relevant axioms for $R$ and $C$\footnote{The
         axioms for real and complex numbers are respectively those
         describing a complete, ordered field \cite{Randolph} and an
         algebraically closed field of characteristic $0$
         \cite{Shoenfield}.}. This assumption is based on the
         prevalent view taken by physics so far of the nature of
         real and complex numbers, that they are elementary
         objects. The only properties of these objects that
         physics cares about are those derivable from the
         relevant axioms.

         The quantum theory representations of real and complex
         numbers described here suggest that $O_{R}$  sees
         that $R_{k,g}$ and $C_{k,g},$ as equivalence classes of Cauchy
         sequence of states of $q_{k}$ strings, represent real and
         complex numbers. $O_{R}$ also sees that $R_{k,g}$ and $C_{k,g},$
         can serve as the base of a frame $F_{k,g}$ containing
         representations of physical theories as mathematical
         structures based on $R_{k,g},C_{k,g}.$

         Symmetry considerations suggest that an observer $O_{k,g},$
         in each frame, $F_{k,g},$ has the same view relative to
         $F_{k,g}$ as $O_{R}$ does relative to the frame $F_{R,C}.$
         Thus $O_{k,g}$ sees $R_{k,g},C_{k,g}$ as elementary,
         structures whose only relevant properties are those
         derivable from the relevant axiom sets. The structure of
         the elements of $R_{k,g},C_{k,g},$ as equivalence classes of
         Cauchy sequences, seen by $O_{R}$, are not visible to
         $O_{k,g}$. Also the construction, in $F_{R,C},$ of representations,
         $R_{k,g},C_{k,g},$ can be repeated in $F_{k,g}$ to obtain
         representations $R_{2,k',g'},C_{2,k',g'}$. Here $2$ is the
         iteration stage. This is visible to an observer $O_{k,g}$ in
         $F_{k,g}.$ The construction in $F_{k,g}$ is possible because
         $F_{k,g}$ contains representations of physical theories,
         including quantum theory, as structures based on $R_{k,g},C_{k,g}.$

         It follows that this construction can be iterated to
         obtain frames emanating from frames.  The iteration or stage
         number provides a third dimension to the frame field where
         for each number $j,$ $F_{j,k,g}$ denotes a frame at stage $j$.

\begin{figure}[ht]\begin{center}
           \resizebox{200pt}{200pt}{\includegraphics[200pt,100pt]
           [660pt,560pt]{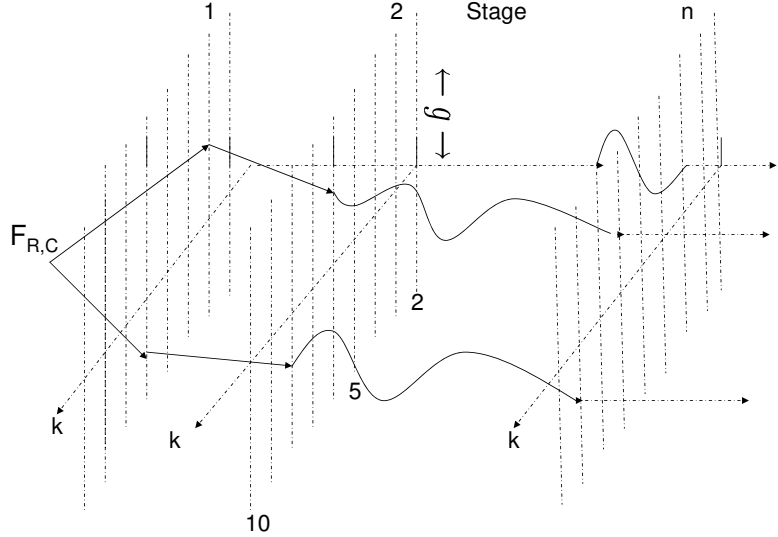}}\end{center}
           \caption{Schematic illustration of a finite number $n$ of
           frame generations coming from frame $F_{R,C}.$ The stage
           number is given at the top. The direction of the iterations
           are shown by the solid connected arrows showing sample iteration
           paths from $F_{R,C}$ through the $k,g$ planes and ending at the
           $nth$ plane. The horizontal dashed lines at the right end
           indicate that in the case of a one-way infinite number of
           iterations there is no terminal stage $n$ for any finite $n$.
           See Fig. \ref{FIQ1} caption for more details.}
           \label{FIQ2} \end{figure}

\begin{figure}[ht]\begin{center}
           \resizebox{200pt}{200pt}{\includegraphics[200pt,130pt]
           [660pt,600pt]{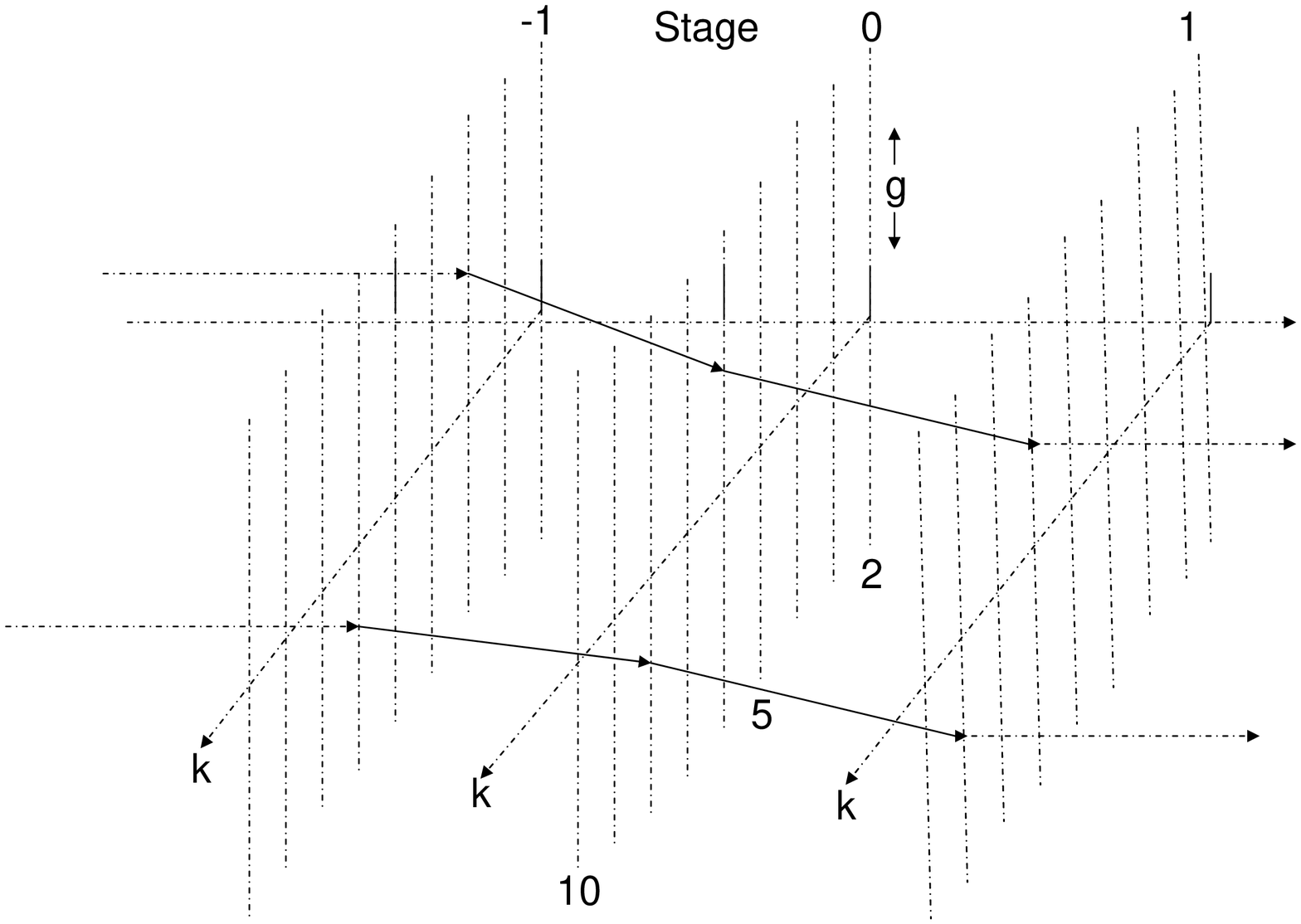}}\end{center}
           \caption{Schematic illustration of a two way infinite number
           of iterations. Here there is no common ancestral frame as
           all frames have parent frames and descendent frames. The
           stage number is given at the top. The direction of the
           iterations are shown by the solid connected arrows showing
           a possible path from one stage to the next with no beginning
           or end.}\label{FIQ3} \end{figure}

\begin{figure}[ht]\begin{center}
           \resizebox{200pt}{200pt}{\includegraphics[100pt,100pt]
           [560pt,520pt]{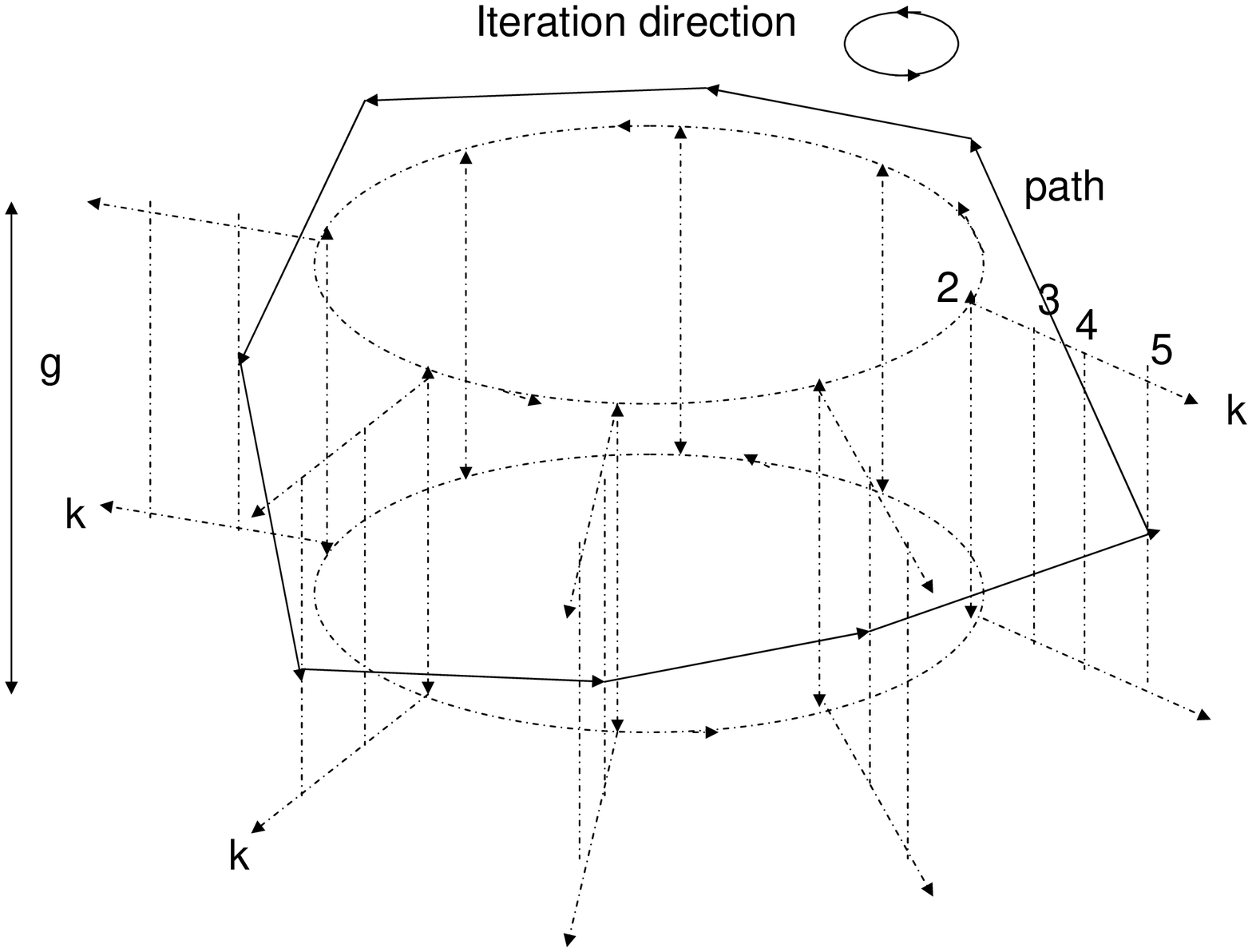}}\end{center}
           \caption{Schematic illustration of a cyclic
           iteration scheme for a finite number (8) of iterations.  There
           is no common ancestral frame as all frames have parent frames
           and descendent frames. The stage number is given at
           the top. The direction of the iterations are shown by the
           solid connected arrows showing a possible cyclic path.
           To avoid clutter, the three $g,k$ planes in the back have
           been suppressed.}\label{FIQ4} \end{figure}

         There are several different iteration types to consider: a
         finite number of iterations, a one way infinite number, a two
         way infinite number, and a finite cyclic iteration. All these
         types are mathematically possible.  They must all be considered
         as there is no a priori reason to choose one type over another.
         The different types are illustrated schematically in figures
         \ref{FIQ2} -\ref{FIQ4}.

           Figure \ref{FIQ2} shows the frame
           field for a finite number, $n,$ of iterations. The iteration
           paths shown represent two out of an infinite number of
           paths. Each path segment, shown by an arrow, stands for a
           quantum theory representation of real and complex numbers
           described in the frame at the arrow tail. The frame at
           the arrow head is based on the described quantum theory
           representation. The iteration direction is shown by the
           arrows.

           Figs. \ref{FIQ1} and \ref{FIQ2} show the existence of a
           fixed frame which is an ancestor for all the frames in
           the field.  This is the case even if $n$ is extended to
           infinity in Fig. \ref{FIQ2} to give a one way infinite
           iteration. Here, too, there is a fixed elementary
           representation of the real and complex numbers that is
           external to the whole field.

           The two way infinite and cyclic iterations shown in Figs.
           \ref{FIQ3} and \ref{FIQ4} are different in this respect.
           There is no representation of the real and complex
           numbers that is external  for the whole
           frame field. All are inside some frame as each frame has
           parent frames. There is no common ancestor frame.


           The path  shown in Figure \ref{FIQ4} for cyclic iterations
           is an example of a path with winding number $1$ in that
           it comes to its starting point in one turn around the
           iteration cylinder. One can, in principle at least have
           paths with  finite winding numbers or even infinite winding
           numbers in that they never return to the starting point.
           One hopes to study in the future these types of paths and
           their dependence on the number of iterations.

           The schematic nature of these figures is to be
           emphasized. Besides showing  that two dimensions of
           the three dimensional frame field are discrete and one,
           the gauge dimension, is continuous, they are very
           useful to show what an observer sees in each frame as
           well as to illustrate the relation between frames in
           different generations. They are also illustrations of the different
           iterations that are mathematically possible.  Which of
           the cases is relevant to physics will have to await more work.

            The relations between the observers in different frames,
           described for Fig. \ref{FIQ1}, is easily extended to multiple
           iteration stages shown in the other figures. Observers in each
           frame have in common the property that they can see down the
           field in the direction of the iterations.  That is they can
           see all their descendent frames, but they cannot see any
           ancestor frames. They also cannot see any other frame at the
           same iteration stage. By "see frames and their relations"
           is meant that  an observer $O_{j,k,g}$ in  frame $F_{j,k,g}$
           can show the presence of the $2$ dimensional frame field
           $F_{j+1}:\{k,g\} \rightarrow \{F_{j+1,k,g}\}.$  This is what
           most of this paper has shown. $O_{j,k,g}$ can also shift the
           derivation by one or more iteration stages to stage $j+2,j+3$
           frames, etc. $O_{j,k,g}$ can also see
           that the $R,C$ representations in these descendent frames
           have structure as equivalence classes of Cauchy sequences
           of (pairs of, for $C$) finite $q_{k}$ string states.

           However $O_{j,k,g}$ cannot describe either ancestor frames
           or other stage $j$ frames. Doing so requires awareness of
           the real and complex number base of a parent frame. These
           are not available as they are outside of $F_{j,k,g}.$

           It is also clear that no observer in a frame can see the
           whole frame field. This view is reserved for an observer
           outside the whole field.\footnote{Here it is assumed that
           any reader of this paper is outside the whole frame field.
           Whether this needs to be revised or not must await further
           work.} An observer, $O_{R,C},$ in a common ancestor frame
           $F_{R,C}$ can see  the whole descendent frame field structure.
           However $O_{R,C}$ cannot tell if there are one or more
           ancestor stages above.

           In many ways this is like the bird (outside the system) and frog
           (inside the system) views used \cite{Tegmark1,Tegmark2} by
           Tegmark\footnote{These concepts
           also play a role in mathematical logic in discussions of
           "absoluteness", i.e. whether or not properties of systems
           in a model of a set of axioms are preserved when one moves from a view
           inside the model to one outside the model.\cite{Jech}}
           In effect one has here a hierarchy of bird and frog views. An
           observer, $O_{j,k,g},$ in a frame, $F_{j,k,g}$ has a frog
           view of $F_{j,k,g}$  and of the theories in  $F_{j,k,g}$.
           $O_{j,k,g}$ sees the real and complex number base, $R_{j,k,g},
            C_{j,k,g}$ as  elementary.  The only relevant properties
            they have are those derived from the relevant set of axioms.
            However, $O_{j,k,g}$ has a birds view of all descendent frames
            in that the relations between all descendent frames are
            visible.

           Cyclic frame iterations present a different situation in
           that descendant frames are also ancestor frames. Because of
           this one may have to relax the stipulation that an observer
           cannot see an ancestor frame. Details of exactly how this
           would occur are not known at present.

           The iteration paths illustrated in the figures give a
           good representation of what observers in different frames
           can and cannot see. Each path is a "visibility" path for
           each frame on the path. If $F_{j,k,g}$ is on a path,
           then any frame $F_{j',k',g'}$ with $j'>j$ on the path is
           a descendent frame and is visible from $F_{j,k,g}$.
           Frames $F_{j',k',g'}$ with $j'<j$ are not visible from
           $F_{j,k,g}$ (except possibly in the cyclic case). The
           totality of frame visibility from $F_{j,k,g}$ is then
           given by the descendant frames in all paths passing
           through $F_{j,k,g}.$

         The presence of a three dimensional frame field shows that
         quantum theory representations have two additional dimensions
         for the frame field that are not present in classical
         representations based on kit strings. One is the presence
         of the freedom of gauge or basis choice.  The other is
         based on the fact that quantum theory, in common with other
         physical theories, is a theory based on the real and complex
         numbers. The relevant point here is that states of finite
         strings of qukits are elements of a Fock space which is
         itself based on the real and complex numbers.  This also
         applies to the states of individual $q_{k}$ which are
         elements of a $k$ dimensional Hilbert space. Both these
         spaces are vector spaces over the complex field $C$.

           \section{Integration with Physics}\label{IP}

           The main problem confronting this work is how to
           integrate quantum representations of real and complex
           numbers and  fields of iterated reference frames with
           physics.  This relationship would be expected to be an
           important part of any approach to a coherent theory of
           physics and mathematics \cite{BenTCTPM,BenTCTPMTEC},
           or to any general theory in which physical and mathematical
           systems are closely related \cite{Tegmark1}. In
           particular one may hope that elucidation of this
           relationship will provide a good foundation to
           theoretical physics. It also may help to decide
           which of the different competing theories of quantum gravity,
           such as loop quantum gravity \cite{Ashtekar} and
           string theory \cite{Zweibach}, is correct.

           \subsection{Simple Relations to Physics}

           There are some simple ways the work
           presented here is related to physics. They are called
           simple only because it is not clear if they would influence
           the properties of physical theories or affect physics.

           One  of these, which was noted earlier, is that  the
           choice of number representations as states of single
           finite $q_{k}$ strings is based on the universality of
           quantum mechanics as a description of physical systems.
           Also influencing the choice is the fact that all physical
           representations of numbers are as states of finite
           strings of physical systems.

           The important role that real and complex numbers have in
           physics should also be stressed. All theoretical predictions
           of physical properties of systems are in the form of real
           numbers as values of physical properties. Also dimensionless
           constants are presumed to be real numbers. Complex
           numbers occur as expansion coefficients of superposition
           states of physical systems and as elements in matrix
           representations of operators.

           Translation of this into the frame field  described here
           has consequences of how the numbers used by physical
           theories in a frame are seen by observers in different
           frames. For example an observer $O_{j,k,g}$ in frame
           $F_{j,k,g}$ sees the real and complex numbers,
           $R_{j,k,g},C_{j,k,g},$ as external featureless objects
           with no properties other than those derived from the real
           and complex number axioms. Any other properties they may
           have are not visible to $O_{j,k,g}.$

           It follows that, from $O_{j,k,g}'s$ viewpoint, all values
           of physical quantities described
           or predicted by physical theory representations, as
           mathematical structures based on $R_{j,k,g},C_{j,k,g}$
           have the same property to $O_{j,k,g}$. This
           applies to both dimensionless physical quantities such as
           the fine structure constant and dimensioned quantities such as
           values of spatial position, distance, momentum, energy,
           all elements of the spectrum of observables,  values of
           the metric tensor $g_{\mu,\nu}$, etc.

           However  an observer $O_{j',k',g'}$ in a parent frame
           $F_{j',k',g'}$ where $j' = j-1$ sees all elements of
           $R_{j,k,g}$ as equivalence classes of states of finite
           $q_{k}$ strings. As a result $O_{j',k',g'}$ also sees
           that all physical quantities described by theories in
           $F_{j,k,g}$ are equivalence classes of Cauchy sequences
           of $q_{k}$ string states.  To summarize, what $O_{j,k,g}$
           sees as elementary featureless objects, are seen by
           $O_{j',k',g'}$ as equivalence classes of Cauchy
           sequences of states of $q_{k}$ strings.

           The same holds for representations of all
           complex valued physical quantities, such
           as elements of matrices representing physical
           transformations, and quantum state expansion
           coefficients. These quantities in frame $F_{j,k,g}$
           are seen by $O_{j',k',g'}$ in $F_{j',k',g'}$
           as equivalence classes of pairs of qukit string states.

           In general, all these results on how the values of physical
           quantities are seen depends on the relation between the
           frame containing the representations of these quantities
           and the viewing frame of an observer. They all follow
           from the observations that in each frame all physical
           theories are represented as mathematical structures based
           on the real and complex number base of the frame.
           How these numbers are seen depends on the relation
           between the frame based on these numbers and the viewing
           frame.

           Because of much interest in quantum gravity and
           associated structure of space and time \cite{QSST}, it is
           worthwhile to consider how a representation of space time
           and its properties in one frame are viewed from a parent
           frame.  As would be expected, real number values of all
           physical properties of space and time, which are featureless
           and elementary in one frame, are viewed as equivalence
           classes of Cauchy sequences of states of finite $q_{k}$
           strings from a parent frame. This applies to distances,
           angles,  coordinate positions, and to values of the
           metric tensor $g_{\mu,\nu}(x)$.  It also applies to
           matrix representations of space time transformations from
           one inertial frame to another.

           In addition, if one regards the points of the space
           time manifold as $4-tuples,$  $R^{4},$ of the real numbers,
           then the same arguments hold. In this case an observer in
           frame $F_{j,k,g}$ sees the points of his own space time
           manifold, $R_{j,k,g}^{4}$ as $4-tuples$ of elementary,
           featureless points whereas an observer $O_{j',k',g'}$ in
           a parent frame $F_{j',k',g'}$ sees the points of $R_{j,k,g}^{4}$
           as $4-tuples$ of equivalence classes of Cauchy
           sequences of $q_{k}$ string states. To $O_{j',k',g'}$ the
           space time points in $F_{j,k,g}$ are not featureless as
           they have structure.

           This  description of how observers describe space time
           representations in different frames is valid only if one describes
           the space time manifold as a $4$-tuple of real numbers.  For
           other descriptions, such as discrete space times or space time foams
           \cite{Hooft,Ng1,Gambini,Sidharth,Hossenfelder,Kempf,Maziashvili}
           or space represented by spin networks \cite{Ashtekar}, it is
           not clear if a similar description applies that is based
           on the relation between the viewing and representation frames.

           Another aspect of integrating the frame field with physics
           is that there is no hint of the frame field structure in
           the properties of the observed physical universe. This suggests
           that one should perhaps find some way to collapse the field
           structure, or at least make the different reference frames
           appear to be "the same" in some sense. This suggests that
           one should require that the physical properties of systems
           represented by frame field elements are frame invariant.
           That is, they are invariant under transformations from one
           frame to another.

           One step in this direction is to require that the field
           structure be such that all frames are equivalent. This
           would restrict the iteration types to the two way
           infinite and finite circular ones as they do not have an
           ancestor frame that is different (from the viewpoint of
           outside the frame field) from the other reference frames.
           It also seems appropriate to restrict consideration to the
           finite cyclic iteration field type, as one way to move
           toward frame invariance is to reduce the size of the frame
           field.

           The ultimate step in this direction is to reduce the
           number of iteration stages in a cycle to just one.
           Whether this is possible or not will have to await future
           work.

           Another approach to reduce the frame field is to eliminate the
           gauge dimension entirely by requiring the states of the
           individual $q_{k}$ to be invariant under any basis change.
           This can be achieved by letting the
           $0$ and $1$ states of each $q_{k}$ be represented by
           different irreducible representations of the gauge group
           $SU(k).$ One method \cite{Enk,Byrd,Kempe} involves constructing new
           qukits from the old ones by reducing the product $SU(k)\times
           SU(k)$ into a sum of irreducible representations and choosing
           any two representations to represent the $0,1$ states of
           each new $q_{k}$.

           Another method \cite{Ashtekar,Mukunda} uses transformations on
           the $SU(k)$  group manifold to construct irreducible representations
           of the group that are invariant under the
           transformations. In essence this is the method used to
           construct angular momentum state subspaces labeled by
           different values of $L$ that are invariant under
           rotations as transformations on $SO(3).$

           \subsection{Speculative Approaches to Integration with
           Physics}
            So far the approaches to integrating the  frame field and
           quantum representations of real and complex numbers with
           physics are rather superficial. They do not represent a
           real integration that treats both physical and
           mathematical systems together in a coherent way.

           How one does this is quite open at present.  However one
           may speculate about various methods to achieve this.  One
           possible way is based on noting that, as units of quantum
           information, the qukits be considered to be fundamental
           objects that can represent either components of
           numbers or physical systems. Whether it represents a
           number component (digit) or a physical system would
           depend on how it is viewed.

           The details of this would have to be worked out to see
           if it has merit.  However, it is worth noting that this
           type of dependence already occurs elsewhere in physics,
           such as the wave-particle and other types of duality.
           Also the suggestion that one consider the prime number
           qukits as elementary and the others as composites, may
           fit in here. In particular the observation is intriguing
           that, if the prime numbers $p$ are related to particle
           spin by $p=2s+1,$ there is just one fermion for $s =1/2;$
           all the rest are bosons.

           Another approach to constructing a coherent integration
           with physics is based on the observation that physical
           theory representations have been inserted into each frame of a
           completed frame field as mathematical structures based on
           the real and complex number base of the frame. Instead
           one may consider involving physical theories in the
           process of constructing Cauchy sequences, their use
           to represent real and complex numbers, and in properties
           of the frame field.

           In this way physical theories may have input into
           constructing their real and complex number bases and,
           conversely, the process of constructing sequences and
           imposing the Cauchy condition may influence the properties
           of the physical theories whose base is being constructed.
           It is even possible that the restrictions imposed by this
           interlocking process may influence the physical
           predictions that the theories can make.

           It would seem that this approach might be most fruitful
           in applying it to the cyclic frame fields and possibly
           those with a very few elements in a cycle. One may
           speculate that the process of closing the cyclic fields
           imposes restrictions on the physical theories and numbers
           involved that influences the values of fundamental constants
           in the theories or predicted values of physical
           quantities.

           Another approach to integrating this work with physics is
           based on the possible representation of a sequence $\Psi$
           as a $\mathcal{B}_{k,g}$ or, more generally, as an $\mathcal{F}_{k}$
           valued quantum field on the nonnegative integers. Then the
           states of the field at each $n$ are given by $\Psi(n).$
           Attention is then restricted to those fields that satisfy
           the Cauchy condition, i.e. the Cauchy fields.

           As was seen, one of the degrees of freedom in representing these
           fields is the freedom of gauge or basis choice. Changes in gauge
           are implemented by gauge transformations acting on the
           fields as shown in Eq. \ref{Umathcal}, or $\Psi'=\mathcal{U}_{k}\Psi.$

           This raises the possibility of using the well developed
           techniques of gauge theories for these fields.  For
           example, one requires that the axioms for the type of
           numbers being described must be invariant under any gauge
           transformation.  Yet it is clear from their expression in
           any particular gauge that their expressions transform covariantly
           under any gauge transformation. The same holds for the
           expression of the Cauchy condition. This is, ultimately,
           a consequence of the gauge dependence shown in Eqs.
           \ref{OpUU} and \ref{ReUU}.

           One should note that the invariance of the axioms of
           number theories under gauge transformations also applies
           to the axioms for any physical theory. The importance of
           this is stems from the fact that all physical
           theories have axioms, whether they are implicit or
           explicitly stated.  Without axioms, theories are empty as
           nothing can be derived or predicted.

           For the gauge theory approach one can ask if there is
           any way to express an action or type of LaGrangian whose
           invariance under gauge transformations expresses the
           invariance of the axioms for numbers and for physical
           theories. If so it may be one way to work towards
           integrating the results obtained here with physics.

           There is an intriguing connection of this approach to the
           standard model in physics \cite{Novaes,Cottingham}. This model is
           a gauge theory where invariance of the LaGrangian under
           gauge transformations requires the introduction of fields
           for the electromagnetic, weak, and strong forces. The
           invariance is under all gauge transformations in the
           group $U(1)\times SU(2)\times SU(3).$

           The connection  of the standard model to the gauge theory
           approach to axiom invariance noted above is based on the
           earlier suggestion  that prime number qukits (those whose
           base $k$ is a  prime number) are elementary and the others
           are composites. Here invariance is under all gauge
           transformations in $U(1)\times SU(2)\times SU(3)\times
           SU(5) \times\cdots.$ The first three groups in the
           product are the same as those for the standard model.
           Whether or not the product of groups $SU(p)$ has to
           include components for all prime numbers or can be cut
           off is not known at present.

           It is not clear if this, or any other speculative approach,
           will work out.  However, these possibilities indicate that
           there is much work needed to integrate  quantum theory
           representations of numbers and the resulting frame field with physics.

           \section{Discussion}
           There are some other aspects of this work that should be
           noted. One is that representation of gauge transformations
           by one continuous dimension of the frame fields, as
           in Figs. \ref{FIQ1}-\ref{FIQ4}, is purely schematic.
           Nothing is implied about what it means for one gauge $g$
           to be close to or far away from another. Indeed it may
           not be useful or even possible to assign a  distance
           measure to the set of gauges.\footnote{Recall that
           each $g$ is a function from $I\times I$ to a basis
           set for a $k$ dimensional Hilbert space associated
           with each element of $I\times I.$}

           In this connection one should note that the choice of
           basis sets $\mathcal{B}_{k,g}$ and $\mathcal{B}_{k',g}$
           in the spaces $\mathcal{F}_{k}$ and $\mathcal{F}_{k'}$ is
           completely arbitrary.\footnote{$\mathcal{B}_{k,g}$
           and $\mathcal{B}_{k',g}$ are each a set of states of all
           finite tuples of states of finite length qukit strings
           for bases $k$ and $k'$.} There is no way to determine if the
           $g$ for the $q_{k'}$ strings is the same or different
           than the $g$ for the $q_{k}$ strings.

           This is different from the usual situation in physics.
           There one has an external reference field or frame that
           can be used to define what it means for a basis of $k$
           dimensional systems to be the same or different from a
           basis of $k'$ dimensional systems. Here no such field
           or common reference frame is present.

           In spite of this the two basis sets can be connected by
           the base changing operator $\tilde{W}_{k',k}$ defined
           earlier. Recall that if the state $|\g,h,s\rangle_{k,g}$
           is in the domain of $\tilde{W}_{k',k},$ then the state
           $|\g,h,s'\rangle_{k,g}=\tilde{W}_{k',k}|\g,h,s\rangle_{k,g}$
           represents the same number in base $k'$ as
           $|\g,h,s\rangle_{k,g}$ does in base $k$.

           This shows that one can proceed in two ways: Arbitrarily
           choose both $\mathcal{B}_{k,g}$ and $\mathcal{B}_{k',g}$
           and define $\tilde{W}_{k',k}$ to be a map from
           $\mathcal{B}_{k,g}$ to $\mathcal{B}_{k',g}.$
           Alternatively choose $\mathcal{B}_{k,g}$ and a
           definition of $\tilde{W}_{k',k}$ and let
           $\mathcal{B}_{k',g}$ be the range set of
           $\tilde{W}_{k',k}.$

           These methods work only if $k$ and $k'$ have the
           same prime factors. If this is not the case, one can
           extend the definition of $\tilde{W}_{k',k}$ by use of
           definitions to accuracy $\ell,$ much as was done for the
           division operator.

           There is another quantum theory representation of real
           and complex numbers that is based on operators instead of
           sequences of states of $q_{k}$ strings.  To define these
           operators one replaces the natural
           number domain of sequences $\Psi$ by states of finite
           $q_{k}$ strings that represent natural numbers.  In this
           way the sequences $\Psi$ become quantum operators $O$ where
           \begin{equation}\label{seqop}\Psi(n)= O|+,h,n\rangle_{k,g}.
           \end{equation} Here $|+,h,n\rangle_{k,g}$ denotes a qukit
           string state $|+,h,s\rangle_{k,g}$ in $\mathcal{B}_{k,g}$
           that represents the number $n$ in base $k$.\footnote{Here
           for simplicity, $|+,h,s\rangle_{k,g}$ is assumed to be
           a state with no leading or trailing $0s.$ This can
           easily be relaxed, if desired.} Note that Eq. \ref{seqop}
           holds irrespective of whether  $\Psi$ is a $\mathcal{B}_{k,g}$
           valued or a more general $\mathcal{F}_{k}$ valued sequence.

           One can use  Eq. \ref{seqop} to replace
           state sequences by operators.  The definition of the
           Cauchy condition can be changed to apply to these
           operators by quantifying over the states
           $|+,h,n\rangle_{k,g}$ as natural number representations
           and replacing the state
           $|\g_{j},h_{j},s_{j}\rangle_{k,g}$ in Eq. \ref{cauchy} by
           $O|+,h,j\rangle_{k,g}$ and the state $\Psi(j)\rangle$
           in Eq. \ref{Pjml}  by $O|+,h,j\rangle_{k,g}.$
           Similar replacements are made for
           $|\g_{m},h_{m},s_{m}\rangle_{k,g}$ and $\Psi(m)\rangle.$
           Operators that satisfy the relevant Cauchy condition are
           denoted here as Cauchy operators.

           The rest of the definition of quantum theory
           representations of real and complex numbers can be taken
           over to define representations as equivalence classes of
           Cauchy operators. In that case there does not seem to
           be a reason why one could not extend the frame field
           description to apply to Cauchy operators. An
           observer in a frame would see real valued physical
           quantities in an immediate descendant frame as
           equivalence classes of Cauchy operators.

           It is clear that there is much to do, both in understanding
           the representations of theories in the frame fields and in
           integrating this work with physics. In any case it is
           seen that quantum representations of real
           and complex numbers as equivalence classes of Cauchy
           sequences of states of qukit strings are different from
           the usual classical representations. Not only are the
           quantum equivalence classes larger than the classical
           ones but the space of representations enjoys two degrees
           of freedom not present in the space of classical
           representations. These are the gauge freedom and the
           iteration stage freedom. The freedom of base choice is
           present in both quantum and classical representations.

           \section*{Acknowledgement}
            This work was supported by the U.S. Department of Energy,
          Office of Nuclear Physics, under Contract No.
          DE-AC02-06CH11357.


\begin{thebibliography}{99}

            \bibitem{Wigner}
          E. Wigner, \textit{Commum. Pure and Applied Math.} {\bf 13} 001
          (1960), Reprinted in E. Wigner, {\it Symmetries and Reflections},
          (Indiana Univ. Press, Bloomington IN 1966), pp222-237.

          \bibitem{Hamming}
          R. W. Hamming, Amer. Mathematical Monthly, \textbf{87},No 2,
          February, (1980).

          \bibitem{Boniolo}
          \emph{The Role of Mathematics in Physical Sciences:
          Interdisciplinary and Philosophical Aspects}, G. Boniolo,
          P. Budinich, M. Trobok, Eds,
          Springer Publications, Dordrecht, the Netherlands, 2005.

          \bibitem{BenTCTPM}
          P. Benioff, Found. Phys. \textbf{32},989-1029,(2002)
          [arxiv:quant-ph/0201093].

          \bibitem{BenTCTPMTEC}
          P. Benioff, Found. Phys. \textbf{35}, 1825-1856,
          (2005)[arxiv:quant-ph/0403209].

          \bibitem{BenRRCNQT}
          P. Benioff, Intl Jour.  Pure, Applied Math. \textbf{39}, 297-341, (2007),
          [arxiv:quant-ph/0508219].

          \bibitem{BenFIQRF}
          P. Benioff,   Jour. Phys. Conference Series,  \textbf{70},012003 (2007),
          [arXiv: quant=ph/0611139]; arXiv: quant-ph/0604135.

          \bibitem{Tegmark1}
          M. Tegmark, arXiv:0704.0646.

          \bibitem{Tegmark2}
          M. Tegmark, Ann. Phys. (NY) \textbf{270}, 1, (1998)
          [gr-qc/9704009].

          \bibitem{Finkelstein}
           D. Finkelstein, Quantum Relativity. Springer-Verlag, Heidelberg
           (1996).

          \bibitem{Schlesinger}
           K.-G. Schlesinger, Journal of Mathematical Physics,
           \textbf{40}, 1344-1358 (1999).

           \bibitem{Titani}
           S. Titani and H. Kozawa, Internat. Jour. Theoret. Phys.
           42, 2575-2602, (2003).

            \bibitem{Takeuti}
          G. Takeuti, \emph{Two Applications of Logic to
          Mathematics} Kano Memorial Lecture 3, Princeton University
          Press, New Jersey, 1978;  \emph{Quantum set theory}, in:
          E. G. Beltrametti, B. C. van Fraassen, Eds.,
           \emph{Current issues in quantum logic},
           Plenum, pp. 303-322, New York 1981.

           \bibitem{Ozawa}
           M. Ozawa, J. Symbolic Logic \textbf{72}, 625-648, (2007),
           [arxiv:math.LO/0604.349].

            \bibitem{Tokuo}
            K. Tokuo, Int. Jour. Theoretical Phys.,
            \textbf{43}, 2461-2481, 2004.

              \bibitem{Corbett}
            J. V. Corbett and T. Durt, arXiv:quant-ph/0211180 v1 2002.

             \bibitem{Davis}
            M. Davis, Internat. Jour. Theoret. Phys.
            \textbf{16},867-874,(1977).

            \bibitem{Gordon}
            E. I. Gordon, Soviet Math. Dokl. \textbf{18}, 1481-1484
            (1977).

            \bibitem{Litvinov}
            G. L. Litvinov, V. P. Maslov, and G. B. Shpiz,
            Archives preprint, quant-ph/9904025, v5, 2002.

            \bibitem{Isham}
            A. Doering and C. J. Isham, arXiv:quant-ph/0703060;
            arXiv:quant-ph/0703062; arXiv:quant-ph/0703064;
            arXiv:quant-ph/0703066.

           \bibitem{Krol}
          J. Krol, "A Model of Spacetime. The Role of
          Interpretations in Some Grothendieck Topoi", preprint,
          (2006).

          \bibitem{Lloyd}
          S. Lloyd, Phys. Rev. Lett. \textbf{88}, 237901, (2002).

          \bibitem{Davies}
          P. C. Davies, arXiv:quant-ph/0703041.

            \bibitem{Hsu}
          S. D. H. Hsu, International J. Modern Phys. A \textbf{22},
          2895-2907, (2007).

          \bibitem{Hewitt}
          E. Hewitt and K. Stromberg, \emph{Real and Abstract
          Analysis} Springer Verlag, Inc.  New York, 1965.


          \bibitem{Burkhill}
          J. C. Burkhill and H. Burkhill, \emph{A second Course in Mathematical
          Analysis}, Cambridge University Press, Great Britain,
          1970.

           \bibitem{Aharonov}
          Y. Aharonov and T. Kaufherr, Phys. Rev. D \textbf{30},
          368-385, (1984).

           \bibitem{Bartlett}
          S. D. Bartlett, T. Rudolph, and R. W. Spekkens, Phys. Rev.
          A \textbf{70}, 032307 (2004).

           \bibitem{BenSQTRN}
          P. Benioff, arXiv:0704.3574 [quant-ph].

          \bibitem{vanEnk}
          S. J. van Enk, Phys. Rev. A \textbf{71}, 032339 (2005).

           \bibitem{Enk}
          S.J. van Enk, Phys.Rev. A \textbf{73}  042306 (2006)
          [arxiv:quant-ph/0602079].

          \bibitem{Poulin}
          D. Poulin and J. Yard, arxiv:quant-ph/0612126.

          \bibitem{Bartlett1}
          S. D. Bartlett, T. Rudolph, and R. W. Spekkens, Rev. Mod. Phys. 79,
          555-609 (2007).

          \bibitem{BenEIPSRN}
          P. Benioff, Phys. Rev. A \textbf{64},052310 (2001)
          [arXiv:quant-ph/0104061v3].

          \bibitem{Barwise}
          J. Barwise, "An Introduction to First Order Logic" in
          Handbook of Mathematical Logic, J. Barwise, Ed.  Studies
          in Logic and the Foundations of Mathematics, Vol. 90.
          North Holland Publishing Co. New york, 1977, pp 5-46.

          \bibitem{Randolph}
          J. F. Randolph, \emph{Basic Real and Abstract Analysis},
          Academic Press, New York, 1968.

           \bibitem{Shoenfield}
            J. R. Shoenfield, \emph{Mathematical Logic}, Addison
            Weseley, Reading, Ma. 1967.

            \bibitem{Jech}
            T.J. Jech, \emph{Lectures in Set theory with Particular
            Emphasis on the Method of Forcing}, Lecture Notes in
            Mathematics No. 217  Springer Verlag New York 1971.

            \bibitem{QSST}
            \emph{The Quantum Structure of Space and Time},
            Proceedings of the 23rd Solvay Conference, Brussels,
            Belgium, Dec. 1-3, 2005, D. Gross, M. Henneaux,
            A. Sevrin, Eds., World Scientific Press, New Jersey.

              \bibitem{Ashtekar}
          A. Ashtekar and J. Lewandowski, Classical and
          Quantum Gravity, \textbf{21}, R53-R152, (2004).

           \bibitem{Zweibach}
          B. Zweibach, \emph{A First Course in String Theory},
          Cambridge Univ. Press, New York, N. Y. 2004.


            \bibitem{Hooft}
            G. 't Hooft, Class. Quant. Grav. \textbf{13} 1023-1040
            (1996) [arXiv:gr-qc/9601014].

            \bibitem{Ng1}
            Y. J. Ng and H. van Dam, Int J. Mod. Phys. A
            \textbf{20},1328-1335, (2005) [arXiv:gr-qc/0403057];
            arXiv:gr-qc/9906003.

            \bibitem{Gambini}
            R. Gambini and J. Pullin, arXiv:gr-qc/0505023.

            \bibitem{Sidharth}
            B. G. Sidharth, arXiv:physics0402007.

            \bibitem{Hossenfelder}
            S. Hossenfelder, arXiv:hep-th/0603032.

            \bibitem{Kempf}
            A. Kempf and R. Martin, arXiv:0708.0062.

            \bibitem{Maziashvili}
            M. Maziashvili, arXiv:0708.1472.

           \bibitem{Byrd}
          M. S. Byrd, D. Lidar, Lian-Ao Wu, and P. Zanardi,
          Phys. Rev A \textbf{71}, 052301 (2005).

          \bibitem{Kempe}
          J. Kempe, D. Bacon, D. A. Lidar, and K. B. Whaley, Phys.
          Rev. A \textbf{63}, 042307 (2001).

          \bibitem{Mukunda}
          N. Mukunda, G. Marmo, A. Zampini, S. Chaturvedi, and R.
          Simon, Jour. Math. Phys. \textbf{46}, 012106 (2005).

          \bibitem{Novaes}
          S. F. Novaes, arXiv:hep-th/0001283;

          \bibitem{Cottingham}
          A. N. Cottingham and D. A. Greenwood, \emph{An Introduction
          to the Standard Model of Physics}, Cambridge University
          Press, Cambridge, UK, 1998.
           \end{thebibliography}
           \end{document}